\documentclass[superscriptaddress,twocolumn,prb]{revtex4-2}
\usepackage{amssymb}
\usepackage{amsmath}
\usepackage{array}
\usepackage{graphicx}
\usepackage{tikz}
\usepackage{pgfplots}
\usepackage{placeins}
\usepackage{relsize}
\usepackage{tabu,multirow}
\usepackage{epsfig}
\usepackage{latexsym}
\usepackage{subfigure}
\usepackage{soul}
\usepackage[export]{adjustbox}
\usepackage{setspace}
\usepackage{float}
\usepackage{stmaryrd}
\usepackage{miller}
\usepackage{bm}
\usepackage{tabularx}
\usepackage{booktabs}
\usepackage[hidelinks]{hyperref} 
\usepackage[noabbrev,capitalize]{cleveref}
\usepackage{ragged2e}

\begin{document}

\title{Atomistic-informed phase field modeling of magnesium twin growth by disconnections}
\author{Yang Hu}
\affiliation{Mechanics \& Materials Lab, Department of Mechanical and Process Engineering, ETH Z\"urich, 8092 Z\"urich, Switzerland}
\author{Dennis M. Kochmann}
\email{Corresponding author}
\affiliation{Mechanics \& Materials Lab, Department of Mechanical and Process Engineering, ETH Z\"urich, 8092 Z\"urich, Switzerland}
\author{Brandon Runnels}
\email{Corresponding author}
\affiliation{Department of Aerospace Engineering, Iowa State University, Ames, IA USA}

\begin{abstract}
The nucleation and propagation of disconnections play an essential role during twin growth. 
Atomistic methods can reveal such small structural features on twin facets and model their motion, yet are limited by the simulation length and time scales. 
Alternatively, mesoscale modeling approaches (such as the phase field method) address these constraints of atomistic simulations and can maintain atomic-level accuracy when integrated with atomic-level information. 
In this work, a phase field model is used to simulate the disconnection-mediated twinning, informed by molecular dynamics (MD) simulations.
This work considers the specific case of the growth of $\hkl{10-12}$ twin in magnesium.
MD simulations are first conducted to obtain the orientation-dependent interface mobility and motion threshold, and to simulate twin embryo growth and collect facet velocities, which can be used for calibrating the continuum model. 
The phase field disconnections model, based on the principle of minimum dissipation potential, provides the theoretical framework. 
This model incorporates a nonconvex grain boundary energy, elasticity and shear coupling, and simulates disconnections as a natural emergence under the elastic driving force. 
The phase field model is further optimized by including the anisotropic interface mobility and motion threshold suggested by MD simulations.
Results agree with MD simulations of twin embryo growth in the aspects of final twin thickness, twin shape, and twin size, as well as the kinetic behavior of twin boundaries and twin tips.
The simulated twin microstructure is also consistent with experimental observations, demonstrating the fidelity of the model. 
\end{abstract}

\maketitle
\section{Introduction}
Twinning is a key deformation mechanism in hexagonal close-packed (HCP) metals and their alloys, which alters the lattice orientation to its mirror image about a twin boundary (TB) and is particularly effective in accommodating strains applied along the $c$-axis.
Certain twin types can be activated at relatively low stresses, such as the $\hkl{10-12}$ tensile twins observed in HCP magnesium (Mg).
These twins are reported to be formed under a loading stress of $\sim$4 MPa (corresponding to a
resolved shear stress of $\sim$2 MPa) in single-crystal Mg \cite{Koike2005,Reed-Hill1957}.
TBs can have both a positive and negative effect on the mechanical properties of HCP metals and alloys. TBs act not only as barriers to dislocation motion but also as nucleation sites for dislocations, facilitating the activation of multiple slip systems.

By engineering the twin structure, it is possible to achieve enhanced mechanical properties of HCP metals and alloys.
A prime example is the gradient-structured AZ31 Mg alloy, which has demonstrated superior performance in strength and ductility compared to its homogeneous, coarse-grained counterpart.
This alloy features a unique structural composition: fine-grained layers adjacent to the surface, layers of parallel twin laminates, and a coarse-grained region densely populated with multi-orientational twins \cite{Zhang2021}.
Another example is the development of gradient twin meshes, such as those consisting of intersecting $\hkl{10-12}$ and $\hkl{10-11}$ twins reported in \cite{Wang2020}, which result in a remarkable increase in ultimate tensile strength and a doubling of ductility.
Further evidence of the advantageous effect of twinning is found in fine-grained AZ31 alloys \cite{Chen2021}, where a high density of $\hkl{10-12}$ twins aligns with increased yield strength and improved elongation to failure.
These examples motivate the need for in-depth knowledge of the twin nucleation and growth mechanisms for designing HCP metals and alloys, emphasizing the intricate relation between twin formation and the overall mechanical performance of HCP metals and alloys. 

Twinning in HCP metals and alloys must be understood as a multiscale mechanism.
At the nanometer scale, small structural features such as disconnections and microfacets on interfaces have been observed \cite{Tu2013,Tu2015,Tu2016,Liu2015,Wang2020characteristic,Wang2022,Zu2017,Sun2014}. 
Disconnections are interfacial dislocations with step characters \cite{Howe2009}, and they have been identified on various interfaces. 
As a representative example, disconnections in $\hkl{10-12}$ twins were reported on basal-prismatic (BP) or prismatic-basal (PB) interfaces and $\hkl{10-12}$ TBs, with interface motion being driven by the nucleation and propagation of these disconnections \cite{Tu2013,Tu2015,Tu2016,Liu2015,Wang2020characteristic,Wang2022,Zu2017,Sun2014,Xu2013,hu2020disconnection}.
The step height of such disconnections is usually one or two interplanar spacings of the BP/PB or TB \cite{Zu2017,Sun2014,Xu2013,hu2020disconnection}, which is a few angstroms. 
The transformation of disconnections formed on BP/PB facets into the twinning disconnections on the twin planes was also found \cite{Zu2017,Sun2014}, and twin embryo growth is mediated by the propagation of twinning dislocations on the primary twin plane \cite{Xu2013,hu2020disconnection}.

The multiscale nature of $\hkl{10-12}$ TBs is evident in their faceted, non-coherent structure, deviating from the ideal $\hkl{10-12}$ twin plane \cite{Tu2013,Tu2015,Tu2016}.
Commonly observed facets on these planes include BP/PB types \cite{Tu2013,Tu2016,Liu2015,Wang2022}, where the basal lattice of the parent/twin aligns with the prismatic lattice of the twin/parent, respectively.
Prior works have revealed that those facets vary in length \cite{Huang2021} but are all of nanometer size.
Short, coherent BP/PB interfaces are highly strained but possess low interfacial energies, even lower than those of coherent TBs \cite{Kumar2015}.
In contrast, long, incoherent BP/PB interfaces are more relaxed, featuring misfit dislocations on the interfaces and high energies.
More recently, new twin facets parallel to the twinning direction have been identified, extending beyond the typical BP/PB facets \cite{Wang2020characteristic}. 
In addition, for twins nucleated in micrometer-sized grains, there is sufficient room to expand and to reach twin lengths of hundreds of microns \cite{Yu2014,Yu2014cozone,ArulKumar2016,Nie2013}. 
Twins that terminate at grain boundaries (GB) can induce the nucleation of twins in neighboring grains and, subsequently, these twins form twin chains that span over multiple grains \cite{ArulKumar2016}.

Atomistic simulations are often utilized to study the energetics of twin transformation \cite{Ishii2016}, or to track the early stages of twin growth when the twin dimensions are of nanometer sizes and the fast growth occurs within nanoseconds \cite{Xu2013,hu2020disconnection,Hu2020}.
However, such simulations face limitations in reproducing twinning behavior at the length and time scales relevant to real-world engineering applications.
To extend to the mesoscale, crystal plasticity (CP) theory has been used to predict twinning and detwinning behavior in materials with HCP structures \cite{Cheng2015,Hollenweger2022,Chang2017}.
Despite its efficiency in modeling the material response on the meso- and macroscopic levels, crystal plasticity struggles to accurately represent the twinning process, primarily because it treats twinning as a unidirectional shear deformation, which oversimplifies the phenomenon \cite{Liu2018}.
Moreover, conventional crystal plasticity models fail to explore the impact of twin structures on the mechanical properties of materials at the nanoscale \cite{kondo2014phase}.

The phase field approach has proven effective in simulating the evolution of twin microstructures without the scale limitations of MD or the kinematic restrictions of CP..
Pioneering work in this area includes the model of \textcite{Clayton2011}, which aimed at identifying the equilibrium twin structure at 0~K, but did not account for issues related to dissipation, growth kinetics, or stress waves.
\textcite{Agrawal2015} later introduced a model with regularized interfaces, which allows for a transparent prescription of interface nucleation and kinetics, allowing for the calibration of phase field models to experimental data or molecular dynamics (MD) simulations.

MD-informed phase field methods have recently accelerated in scope and accuracy.
For example, the phase field model proposed by \textcite{Amirian2022} uses the computed $\hkl{10-12}$ TB energy by density functional theory (DFT); this model was calibrated with the twin tip and TB velocities from MD \cite{hu2020disconnection}.
While the model was able to obtain the kinetic coefficient (mobility parameter) for interface motion, it did not capture the faceting of TBs.
Alternatively, \textcite{spearot2020structure} and  \textcite{gong2021effects} proposed phase field models that examine the motion of disconnections during twin growth in Mg. 
Even though informed by atomistic simulations or anisotropic interface energy and mobility, the use of linear kinetics reduced the accuracy for prediction of individual facet mobility. 
Moreover, without the inclusion of an explicit regularization scheme, possibly compromising the effects of curvature and anisotropy.

The phenomenon of disconnection-mediated twin growth falls under the broad category of microfacetting and disconnection migration, which is common to many GBs (e.g.~\cite{Fang2022,Zhu2019,Sternlicht2019,Sternlicht2022}) and has been explained and modeled in a continuum mechanical setting by \textcite{Chesser2020}.
These models explicitly incorporate pronounced nonconvexity in GB energy, elasticity, shear coupling, and the principle of minimum dissipation potential for GB motion, being capable of capturing the motion of disconnections on GBs without prior knowledge of their existence. 
Importantly, they also include nonlinear kinetics (via a thresholding method), so that interface moves only after the driving force exceeds the motion threshold.
This creates a profound difference from traditional linear kinetics (as also reported in other phase field settings \cite{Guin2023}) and yields results that are impervious to the mesh.
Building on this framework, \citet{Gokuli2021} included a nucleation model for simulating the thermal nucleation of disconnection pairs.
Although being primarily applied to bicrystals for studying the migration of individual GBs, with minor modifications, the models reported in \cite{Chesser2020,Gokuli2021} are suitable for simulating twin growth in terms of both lengthening and thickening.

In this work, the phase field disconnections model (\cite{Chesser2020,Gokuli2021}) is tailored to the problem of Mg twin growth by including MD-informed information, with full anisotropy, to predict twin facet motion in twin embryo growth at larger length and time scales than what is accessible by MD. 
The phase field model is optimized by including a nonlinear kinetic relation for interface motion and an orientation-dependent motion threshold suggested by MD simulations.
Such integration of MD results into a phase field model effectively bridges the gap in length and time scales between atomistic and mesoscopic simulations, while preserving the mechanistic insights into the deformation process.
The growth of the $\hkl{10-12}$ twin in Mg is chosen as the subject of interest, given the recent surge in interest in Mg due to its light weight and high specific strength.
As explained above, $\hkl{10-12}$ twins are particularly profuse during the plastic deformation of Mg.

The remainder of this work is organized as follows: \cref{sec:md_simulations} covers the computational details of the MD simulations and results; \cref{sec:phase_field_modeling} presents the detailed description of the phase field model and results; \cref{sec:discussion} discusses the power and limitation of the developed phase field model.
Conclusions are summarized in \cref{sec:conclusion}.

\section{Atomistic simulations}\label{sec:md_simulations}
MD simulations are performed in the Large-scale Atomic/Molecular Massively Parallel 
Simulator (LAMMPS) \cite{Plimpton1995} with a modified embedded atom method (MEAM) potential proposed in \cite{Kim2009}.
Two types of simulations are performed. First the motion of individual facets in bicrystals is simulated to obtain the interface mobility and the threshold driving force for interface motion.
Second, the growth of a single twin embryo is simulated, whose results are used to calibrate the developed phase field model. 
 
\subsection{Potential validation}
\begin{figure*}
  \centering
  \includegraphics[width=0.9\textwidth,clip,trim=3.7cm 1cm 2cm 0cm]{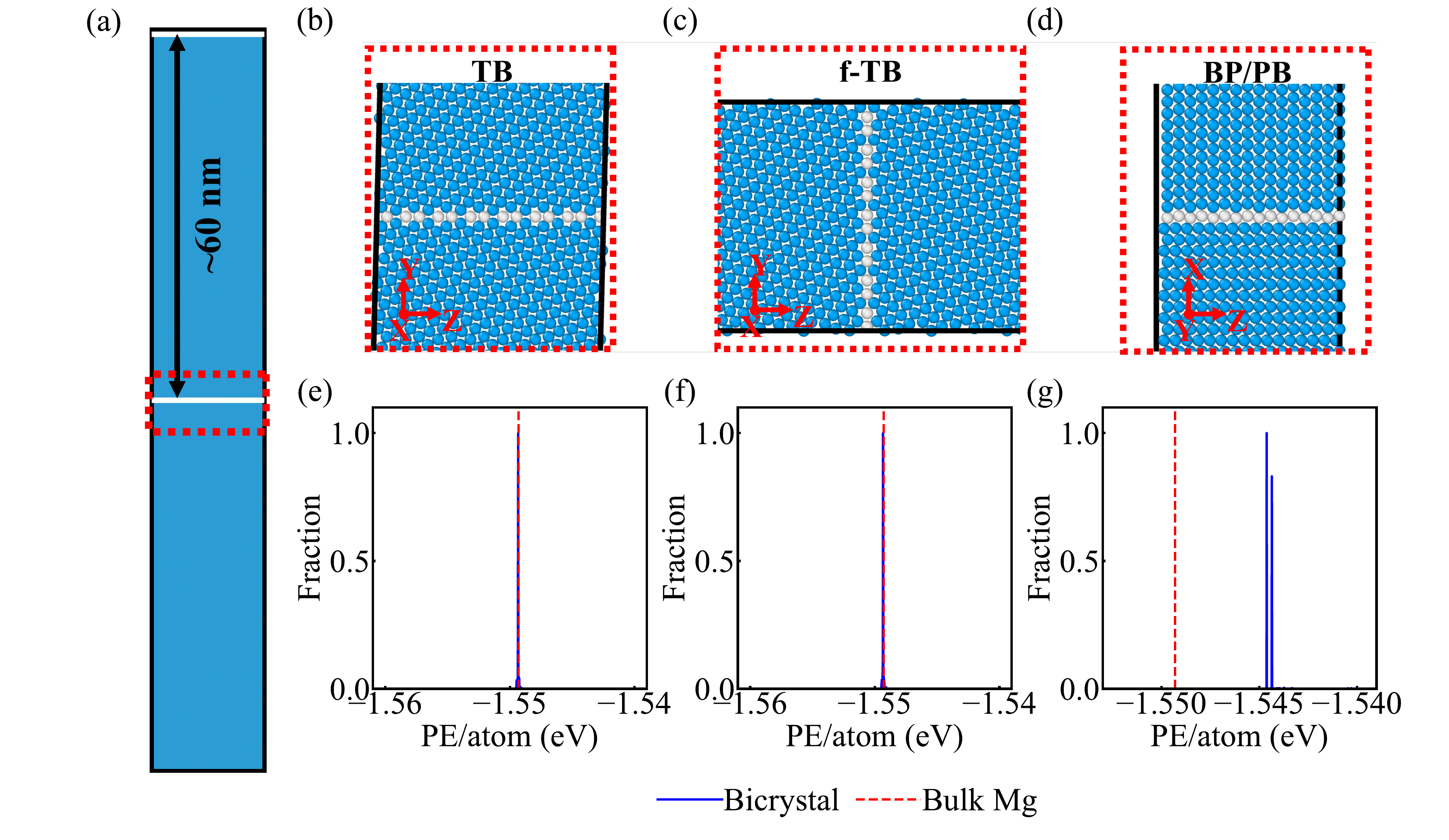}
  \caption{(a) The MD simulation box for simulating individual facet motion, with zoomed-in views of the atomic structures of (b) TB, (c) f-TB, and (d) BP/PB interfaces and the corresponding potential energy (PE) value per atom distribution for (c) TB, (f) f-TB, and (g) BP/PB interfaces.}
  \label{Figure 1}
\end{figure*}

The potential used for simulations is first validated by calculating various material parameters and comparing them with existing data.
These parameters, including the interfacial energies of TB and BP/PB interfaces as well as elastic constants of Mg, are closely linked to the twinning behavior of Mg \cite{hu2023atomistic}.
In addition, the calculated values will be fed into the developed phase field model.
Interfacial energies are calculated as
\begin{equation}
\gamma = \frac{E_\text{int}-n \cdot E_\text{atom}^\text{bulk}}{A},
\end{equation}
where $A$ is the interface area, $E_\text{int}$ is the potential energy of the structure containing two interfaces, and $E_\text{atom}^\text{bulk}$ is the potential energy of a single atom far from the interfaces (see \cref{Figure 1}).
For fully relaxed interfaces, this latter value is the same as that calculated from a defect-free, bulk sample.
Yet, for non-relaxed interfaces, the lattices on the two sides of the interface are strained, so that the atomic potential energy differs from the value obtained from a fully relaxed bulk sample.
The calculated interfacial energies, using the MEAM potential of \cite{Kim2009}, are shown in \cref{Table 1} together with reported values from \cite{Kumar2015,Huang2021Hydrogen,Wang2012}.
The elastic constants of HCP Mg at 0~K were calculated as the coefficients that relate stresses and strains.
The computed results are shown in \cref{Table 2} together with reference values from \cite{Jain2013,Yang2021,Slutsky1957}.
Overall, there is reasonable agreement between the values calculated using the MEAM potential and literature data, so that the chosen potential is deemed a reasonable approximation to describe twinning in Mg.

\begin{table}
  \caption{Energies of coherent TB and BP/PB interfaces}
  \label{Table 1}
  \begin{tabularx}{\linewidth}{XXl}
    \toprule
    Interface & \multicolumn{2}{l}{Interfacial energy ($mJ$/$m^{2}$)}   \\ 
    & Here & Literature   \\ 
    \midrule
    TB & 143.9 & 135.1 \cite{Kumar2015}, 130.2 \cite{Huang2021Hydrogen}, 122.3 \cite{Wang2012}   \\ 
    BP/PB & 106.7 & 101.0 \cite{Kumar2015} \\
    \bottomrule
  \end{tabularx}
    \justify{\footnotesize\it Note:
    The TB and BP/PB energies reported in \cite{Kumar2015} were calculated by density functional theory (DFT), using the projector augmented wave (PAW) approach and the Perdew-Burke-Ernzerhof (PBE) generalized gradient approximation (GGA) for exchange-correlation.
    The TB energy from \cite{Huang2021Hydrogen} was calculated using a similar method. The TB energy reported in \cite{Wang2012} was obtained using the EAM potential of \cite{Liu1996}.}
\end{table}

\begin{table}
  \caption{Magnesium elastic constants}
  \label{Table 2}
  \begin{tabularx}{\linewidth}{XXl}
    \toprule
    Component & \multicolumn{2}{l}{Modulus value (GPa)} \\
    & Here & Literature \\ 
    \midrule
    $C_{11}$ & 62.8 & 76 \cite{Jain2013}, 63.96 \cite{Yang2021}, 63.5 \cite{Slutsky1957} \\ 
    $C_{33}$ & 69.6 & 71 \cite{Jain2013}, 67.27 \cite{Yang2021}, 66.5 \cite{Slutsky1957} \\ 
    $C_{44}$ & 17.1 & 19 \cite{Jain2013}, 17.02 \cite{Yang2021}, 18.4 \cite{Slutsky1957} \\
    $C_{66}$ & 18.4 & 31 \cite{Jain2013}, 18.7 \cite{Yang2021} \\
    $C_{12}$ & 26.0 & 15 \cite{Jain2013}, 28.30 \cite{Yang2021}, 25.9 \cite{Slutsky1957} \\
    $C_{13}$ & 21.2 & 19 \cite{Jain2013}, 20.51 \cite{Yang2021}, 21.7 \cite{Slutsky1957} \\
    \bottomrule
  \end{tabularx}
  
  \justify{\footnotesize\it
    Note: The elastic constants from \cite{Jain2013,Yang2021} were calculated by DFT using the PAW approach and the PBE-GGA exchange-correlation functional. 
    The values reported in \cite{Slutsky1957} are 0~K extrapolations of adiabatic elastic constants of Mg single crystals, measured by an ultrasonic pulse technique at various temperatures.
  }
\end{table}

\subsection{Atomistic simulation setup of individual twin facet motion}
To simulate the motion of individual facets, MD simulation boxes containing two interfaces are constructed with periodic boundary conditions along all axes being maintained (\cref{Figure 1}a).
The distance between two interfaces is over 60~nm to avoid strong interface interactions.
For the box containing two TBs, the $X$-axis is set to be parallel to the $\hkl[12-10]$-direction, and the $Y$-axis is along the twinning direction ($\hkl[10-11]$-direction), so that the $XY$-planes are parallel to the TB (\cref{Figure 1}b-c). The cross-sectional area is around  $5.1 \times 5.3\,\text{nm}^{2}$.
For the box containing two forward twin boundaries (referred as f-TBs, which are under $\sim90^\circ$ from the horizontal TBs, also known as conjugate twin boundaries), the same box orientation is used, while the $XZ$-plane is parallel to the f-TB with a cross-sectional area of around $5.1 \times 5.3\,\text{nm}^{2}$.
For boxes with BP/PBs, the $Y$-axis is parallel to the $\hkl[12-10]$-direction, and the $X$-axis is set along the $\hkl[-1010]$-direction, so that the $XY$-plane is parallel to the BP/PB plane (\cref{Figure 1}d).
The cross-sectional area is about $4.3 \times 4.5\,\text{nm}^{2}$.
Here, BP interfaces are those between the basal lattice (as the twinned region) and the prismatic lattice (as the parent), while the prismatic lattice serves as the twinned region for PB interfaces.
Among all facets, TBs and f-TBs are fully relaxed, meaning that the atoms in regions far from the interfaces have the same potential energies as those in a defect-free bulk sample (\cref{Figure 1}e-f).
By contrast, coherent BP/PBs only exist with the two lattices above and below being strained, due to the mismatch of the differently spaced basal and prismatic planes at the interface.
Compared to unstrained basal and prismatic lattices, the strain tensor applied to the basal part in order to form a coherent interface is $\operatorname{diag}(-0.034,0.004,0.009)$, and the strains applied to the prismatic part are $\operatorname{diag}(0.033, 0.004, -0.011)$. This strain increases the potential energy of atoms on both sides of the interface (\cref{Figure 1}g).
Introducing misfit dislocations can relax the stress but increases the interfacial energies, see \cite{Ishii2016,Xu2013}. 

To activate the motion of TBs and f-TBs, constant shear strains are applied by uniformly shearing the simulation box along the $Y$-axis. Positive (negative) applied strains imply deforming the box along the positive (negative) $Y$-axis. Atoms are relaxed under an NVT (isothermal-isochoric) ensemble at 1~K, using the Nos\'{e}-Hoover thermostat, with the temperature adjusted every 100 time steps with one integration step of 0.1~fs. BP/PBs are usually $\sim$45$^\circ$ or $\sim$135$^\circ$ away from TBs about the $a$-axis, so that shear strains applied parallel to the TB cause tension along the plane normal direction and a compression along an axis that is parallel to the plane, and vice versa.
To mimic the same loading condition as in the simulation with TBs and f-TBs, normal strains with opposite signs are applied to the $Y$- and $Z$-axes to simulation boxes with BP/PBs.
In this work, if the strains applied along the $Z$-axis are tensile strains, they are considered positive.

To differentiate the twinned region from the parent, structural analysis and visualization of atomic configurations are performed using the open-source visualization tool OVITO \cite{Stukowski2010}. The Polyhedral Template Matching (PTM) method \cite{Larsen2016} is used to characterize the local crystalline structure and orientation associated with each atom in the system, which can be encoded as an orientation quaternion $\boldsymbol{q} = \boldsymbol{q}_{w} + q_{x}\boldsymbol{i} + q_{y}\boldsymbol{j} + q_{z}\boldsymbol{k}$.
Components $q_{x}$ and $q_{w}$ present suitable criteria to identify the twinned region from the parent (\cref{Figure 2}a, 2d and 2g).

\subsection{Calculating the driving force for facet motion and facet mobility} \label{sec:calculating_driving_force}

\begin{figure*}[ht]
  \centering
  \includegraphics[width=0.9\textwidth,clip,trim=0cm 2.5cm 2cm 1cm]{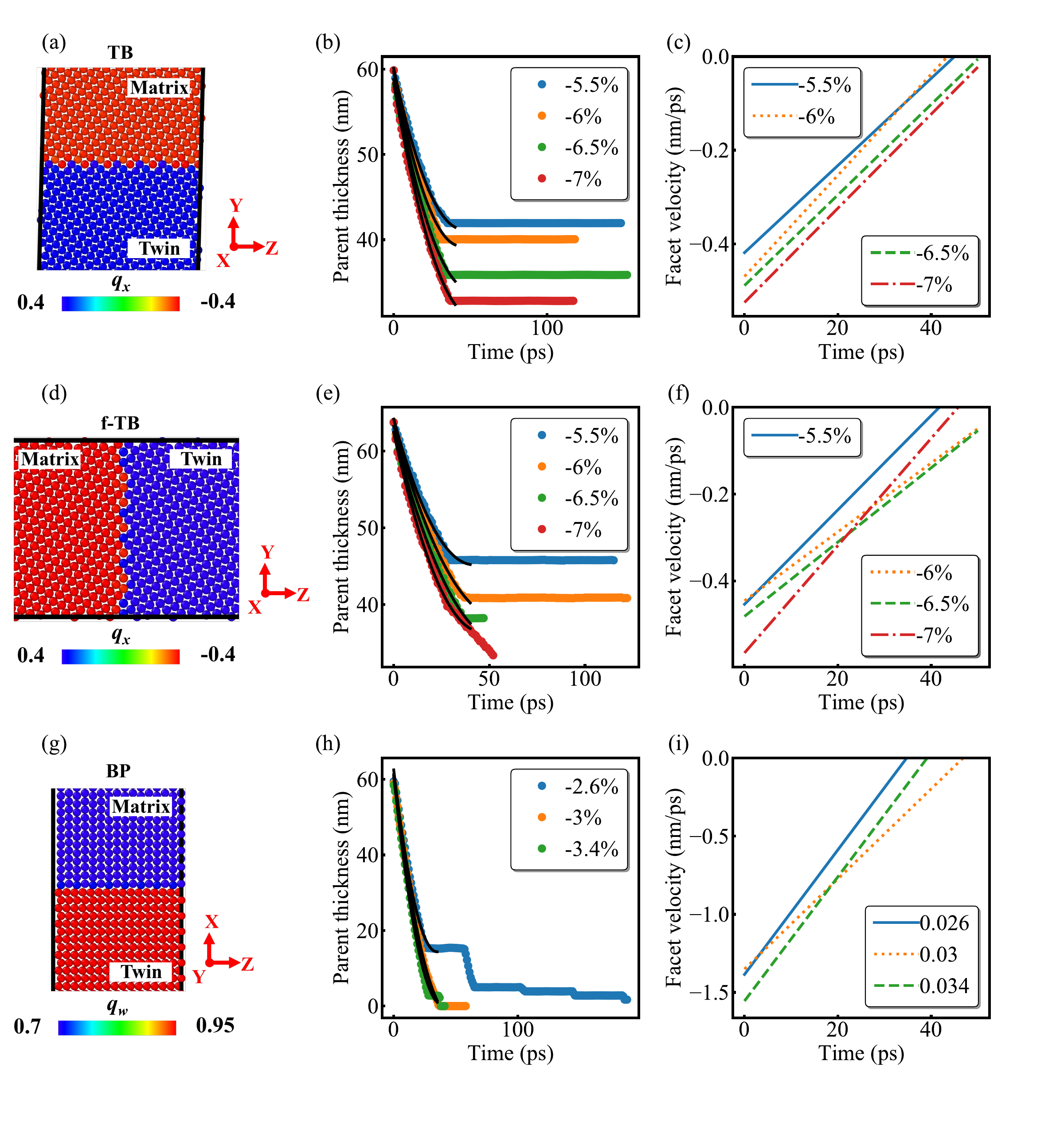}
  \caption{
    (a) Magnified view of the atomic structure of a TB.
    The time evolution of (b) the thickness of the parent and (c) facet velocity for simulations with two TBs.
    (d) Magnified view of the atomic structure of a f-TB.
    The time evolution of (e) the thickness of the parent and (f) the facet velocity for simulations with two f-TBs.
    (g) Magnified view of the atomic structure of a BP/PB interface.
    The time evolution of (h) the thickness of the parent and (i) facet velocity for simulations with two BP interfaces.
    In (a), (d) and (g), atoms are colored by the component $q_{x}$ or $q_{w}$ of the orientation quaternion.  }
  \label{Figure 2}
\end{figure*}

The thicknesses of the parent and/or the twin at different times are measured under applied strains. 
Regardless of the facet type, the parent region shrinks when negative strains are applied, and the region thickness eventually reaches an equilibrium (\cref{Figure 2}b, 2e, and 2h).
The larger the magnitude of the applied strain, the more the twinned region expands and the smaller the final parent thicknesses.
Analogously, under positive applied strains, the parent region grows at the expense of the twinned region.

The velocity of each facet is estimated by first fitting the variation of the thickness of the parent region with time, using the quadratic regression, and then calculating the time derivative of the evolving parent thickness. The initial velocities are the resulting slopes in the initial instance at 0~ps. 
Comparing with the instantaneous velocities (the slope of the thickness-time curve at each time) which fluctuate significantly, the facet velocities estimated using the quadratic regression exhibit a smoother trend with the driving force, which will be shown later. Moreover, they represent an average instantaneous velocity within a certain period of time. For example, for the TB motion at 7\% of shear strain, the average instantaneous velocity within 10, 15, and 20 ps is about 0.572, 0.497, and 0.458 nm/ps, respectively, and the TB velocity at 0 ps estimated by the quadratic regression is about 0.515 nm/ps, being close to the average instantaneous velocity within 15 ps from the beginning of the simulation. 
Facet velocities are considered negative when the parent region shrinks, and vice versa.
It is observed that the facet velocity decreases with time, due to the constant strain condition (\cref{Figure 2}c, 2f and 2i).
In MD simulations, the total energy and potential energy of the simulated sample decrease with time until, given sufficient simulation time, the sample structure will reach its energetically most favorable state and the energies reach an equilibrium. 

To calculate an interface mobility, it is necessary to determine the thermodynamic driving force onto the interface.
For a mechanical interface, the driving force $f$ (also known as driving traction, as it is a force per area) is derived as \cite{Abeyaratne1990,Guin2023}
\begin{equation}
f = \llbracket W \rrbracket - \langle\boldsymbol{\sigma}\rangle  \llbracket \boldsymbol{\varepsilon} \rrbracket + \gamma \kappa
= \llbracket\mathrm{G}\rrbracket + \gamma\kappa
\end{equation}
where $W$ is the Helmholtz free energy density, $\bm{\sigma}$ and $\bm{\varepsilon}$ are the Cauchy stress and infinitesimal strain tensors, respectively, $\mathrm{G}$ is the Gibbs free energy.
The notation $\llbracket\cdot\rrbracket$, $\langle\cdot\rangle$ indicates the difference and average, respectively, in quantity $(\cdot)$ between the twin and parent.
$\gamma \kappa$ is the contribution from the interfacial energy, where $\gamma$ is the interfacial energy and $\kappa$ is the curvature of the interface.
In the case of bicrystals, $\kappa=0$.

The mechanical response is approximated as linear elastic, which is an appropriate simplification when both the twin and parent are close to their respective mechanical equilibrium configuration, so that stresses and strains (about that respective equilibrium) are small. In this case, the strain energy density has the form
\begin{equation}
W 
= \frac{1}{2}\bm{\sigma} : \mathbb{C}^{-1} \bm{\sigma} 
= \frac{1}{2}\bm{\varepsilon}_e:\mathbb{C}\bm{\varepsilon}_e,
\end{equation}
where $\mathbb{C}$ is the fourth order elastic modulus tensor, differing between the parent and the twin, and it is understood that $\boldsymbol{\varepsilon}_e$ is the (infinitesimal) elastic strain tensor about the respective equilibrium lattice configuration in either twin or parent. 
The elastic tensor of the twin, $\mathbb{C}_\text{T}$, is obtained by rotation of the elastic tensor $\mathbb{C}_\text{P}$ of the parent via the transformation
\begin{align}
    (\mathbb{C}_\text{T})_{pqst} = R_{pi}R_{sk}(\mathbb{C}_\text{P})_{ijkl}R_{qj}R_{tl},
\end{align}
where $R_{ij}$ are the components of $\mathbf{R}\in SO(3)$, which is the misorientation matrix (and the summation convention over repeated indices is implied).
$\boldsymbol{\sigma}$ is computed as the virial atomic stresses by LAMMPS; the average atomic stresses of the parent and the twin are obtained by summing the values for each atom in the region, averaged over the volume of the region. 
Analogous to the decreasing trend of facet velocity with time, the atomic stresses also change with time and thus the calculated driving force.
The total strain associated with each atom is estimated using the Atomic Strain modifier embedded in OVITO  \cite{Shimizu2007,Falk1998}, from which averages are computed as before.

\begin{figure}
  \centering
  \includegraphics[width=1\linewidth,clip,trim=2cm 1.7cm 0.5cm 4cm]{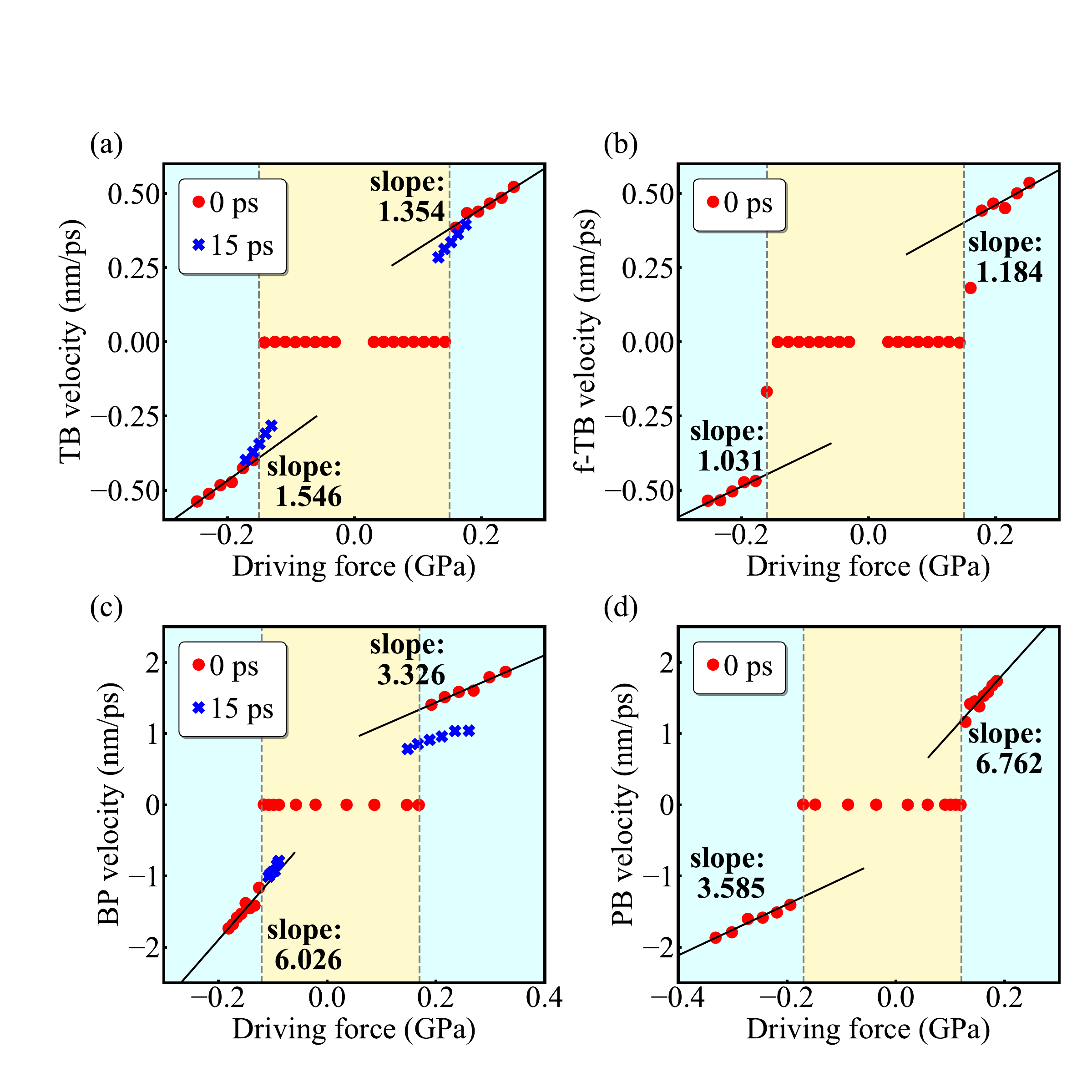} 
  \caption{Variation of facet velocity with the driving force for a (a) TB, (b) f-TB, (c) BP, and (d) PB interface. Red circles indicate interface velocities and driving forces at $t = 0$~ps; blue crosses in (a) and (c) are computed at $t=15$~ps.}
  \label{Figure 3}
\end{figure}

The facet velocities at~0 ps, estimated using the quadratic regression, and the driving forces are calculated at 0~ps (\cref{Figure 3}).
Under negative applied strains, the driving forces for the motion of TBs, f-TBs, and BP interfaces are negative, so the twinned region expands.
For PB interfaces, their direction of motion is opposite to that of BP interfaces.
In other words, if a BP interface moves upward, it is equivalent to a PB interface moving downward, and positive strains along the $Z$-axis lead to twin expansion.
For all four facets, the smaller the driving force (magnitude), the smaller the facet velocity (magnitude), and as the driving force drops below a threshold value, the interface becomes static (its velocity vanishes).

\begin{figure}
  \centering
  \includegraphics[width=\linewidth,clip,trim=0.4cm 0cm 0.6cm 0cm]{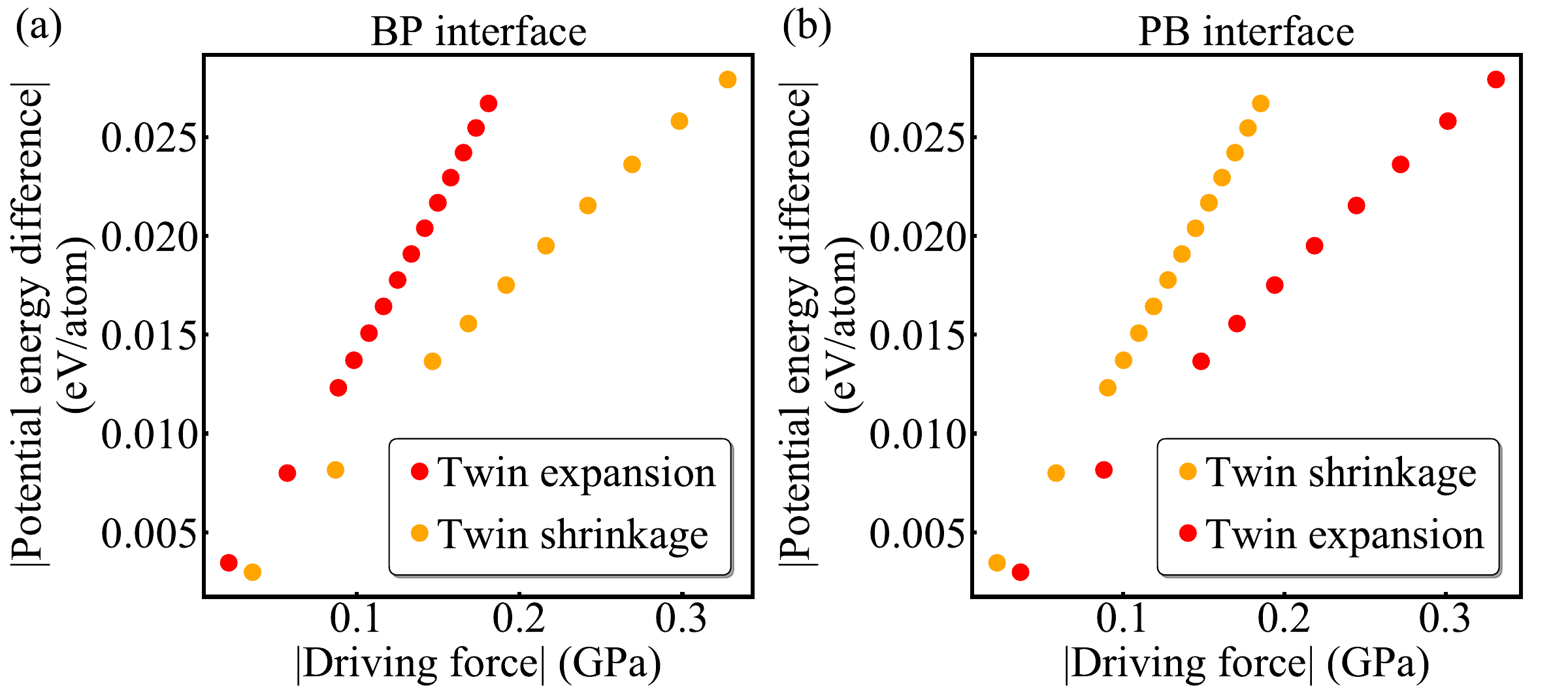}
  \caption{Magnitude of the atomic potential energy difference (a) between the twin and parent during BP interface motion versus the magnitude of the BP interface driving force, (b) between the twin and parent during PB interface motion versus the magnitude of the PB interface driving force.}
  \label{Figure 4}
\end{figure}

The interface mobilities are obtained by fitting the variation of facet velocity with driving force using linear regression.
TB motion under shear strains of opposite signs revealed similar mobilities.
The motion of f-TBs is consistent with the motion of TBs, as expected, since f-TBs have similar atomic structures and interfacial energies to TBs.
The mobilities of BP/PBs are four to five times larger than those of TBs and f-TBs.
The motion of BP/PB also changes under applied strains of opposite signs: at the same magnitude of driving force, the interface velocity is higher when the basal lattice is expanding, meaning that the transformation from the prismatic lattice to the basal lattice is energetically more favorable than the opposite process.
This is further confirmed by considering the difference in magnitude of the atomic potential energy between the basal and prismatic lattice vs. the magnitude of driving forces for the twinning and detwinning processes (\cref{Figure 4}).
For BP interfaces, energy is reduced more during the twin expansion process (i.e., prismatic $\rightarrow$ basal) than the parent expansion (i.e., basal $\rightarrow$ prismatic).
For PB interfaces, the twin expansion process (i.e., basal $\rightarrow$ prismatic) releases less energy.
The different motion of BP/PB interfaces under strains of opposite signs suggests that different interface mobilities may need to be considered when studying twinning and detwinning processes.
The prismatic and basal lattices are strained slightly differently when forming the interface (resulting in only negligible differences in potential energy far from the interface), 
but thorough DFT calculations are necessary to accurately quantify the energetics associated with such transformation for further insight into the asymmetry in migration behavior. 


\subsection{Simulating the growth of a single twin embryo}
\begin{figure}
  \centering
  \includegraphics[width=\linewidth,clip,trim=0cm 0.2cm 0cm 0cm]{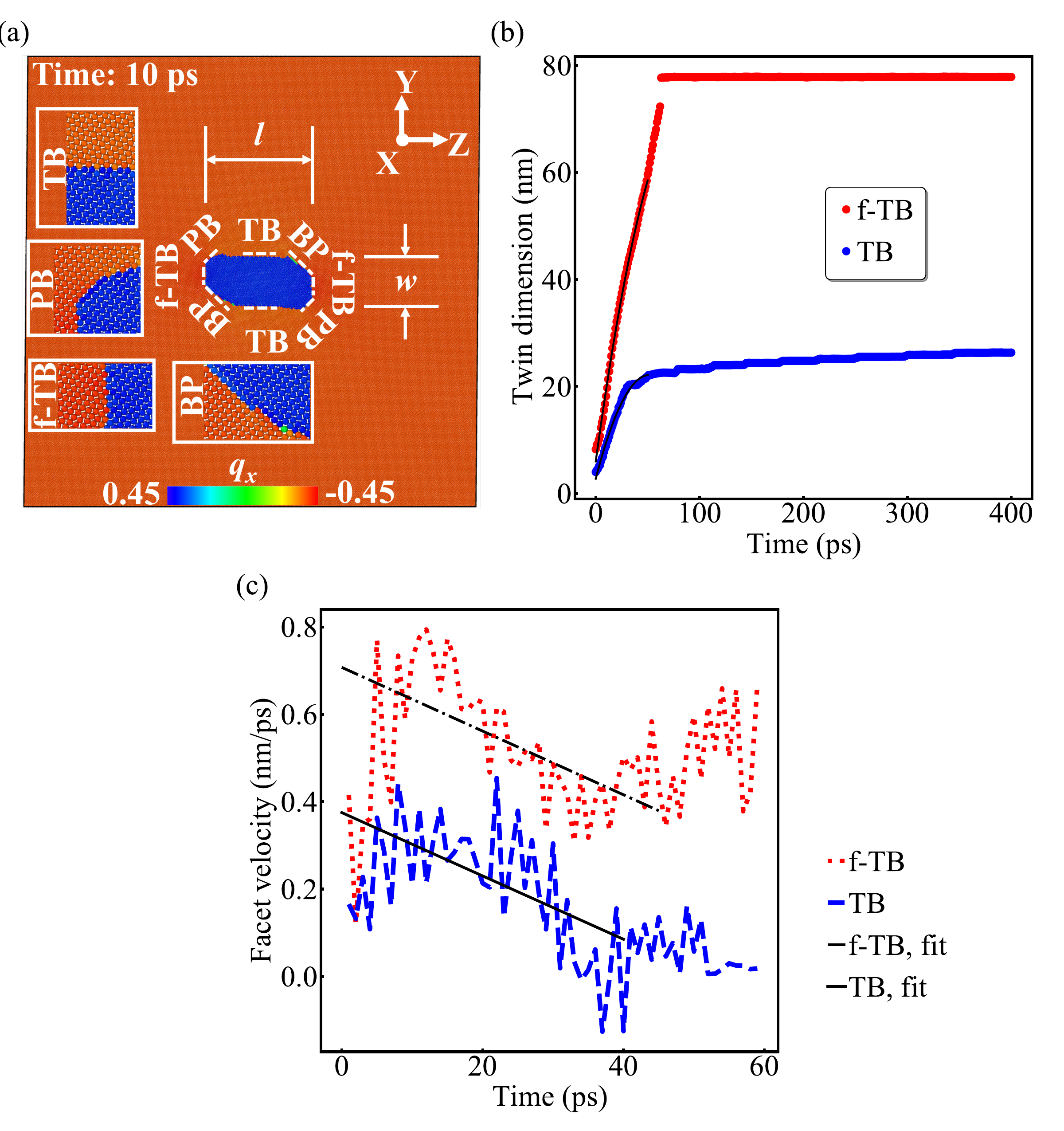}
  \caption{(a) Twin microstructure at 10~ps simulated by MD, with magnified views of TB, f-TB, and BP/PB interface. Atoms are colored according to the component $q_{x}$ of the orientation quaternion. (b) The time evolution of twin thicknesses and lengths. (c) The time evolution of facet velocity for TBs and f-TBs.}
  \label{Figure 5}
\end{figure}

The simulations of twin embryo growth follow the procedures previously introduced in \cite{Hu2020}.
The simulation box has the same orientation as the bicrystal with two TBs, see \cref{Figure 5}a.
The dimensions of the simulation box are about 5.1 $\times$ 77.2 $\times$ 77.0 $\text{nm}^{3}$ in the $X$-, $Y$-, and $Z$-directions, respectively, and the box contains approximately 1.3~million atoms.
A twinned region of a $\hkl{10-12}$ twin with a length of 7.6~nm and a thickness of 4.1~nm is inserted at the center of each simulation box, using the Eshelby method \citep{Huang2021}.
The initial twin structure is obtained after structural relaxation, and the boundary that separates the twin and the parent includes TB, f-TB, and BP/PB interfaces.
The combination of BP/PB and f-TB is also referred to as the twin tip (TT) in this work.
Magnified views (\cref{Figure 5}a) show the atomic structure of each interface type, which are consistent with the interfaces in the bicrystals shown above.
To activate twin growth, a global shear stress of about 1.3~GPa is first applied parallel to the $\hkl{10-12}$ planes ($XY$-plane) in the $\hkl[10-11]$-direction ($Y$-axis), using an NPT (isothermal-isobaric) ensemble at 1~K.
The simulation box is held for 1~ps to reach an equilibrated shear stress.
Subsequently, an NVT (isothermal-isochoric) ensemble is employed to maintain the shear strain value that is responsible for the 1.3~GPa shear stress, and the simulation box is relaxed. 

The twin lengths ($l$) and thicknesses ($w$) are approximated by the differences between the maximum and minimum positions of the twin along the $Y$- and $Z$-axes, respectively.
The velocity of each facet at a specific time is determined by calculating the slope of the time evolution of the twin dimensions.
The twin boundaries move faster along the $Y$-axis than the $Z$-axis (as shown in \cref{Figure 5}b), leading to a larger final twin length than thickness.
This is further confirmed by the calculated facet velocity, which also shows varied facet velocities during twin growth.
There is a slow twin expansion regime within the initial $\sim$2~ps of the simulation, indicating an incubation period before discernible twin growth starts.
A rapid twin growth stage follows, and the facet velocity eventually decreases with time because of the constant shear strain maintained throughout the simulation.
For twin propagation along the $Y$-axis, there is a slight increase in the velocity due to the strong interaction between the twin and its periodic image as the TTs approach the edge of the simulation box near the end of the simulation.
The facet velocities are estimated by quadratic fitting of the twin dimensions over time and taking the resulting slopes in the initial instance at 0~ps (see the slopes plotted in \cref{Figure 5}c).
The velocity at 0~ps estimated by the quadratic regression (0.71~nm/ps for TT and 0.40~nm/ps for TB) is indeed similar to the average of the instantaneous velocities from 5 to 15~ps (0.67~nm/ps for TT and 0.30~nm/ps for TB). 

\begin{figure}
  \centering
  \includegraphics[width=\linewidth]{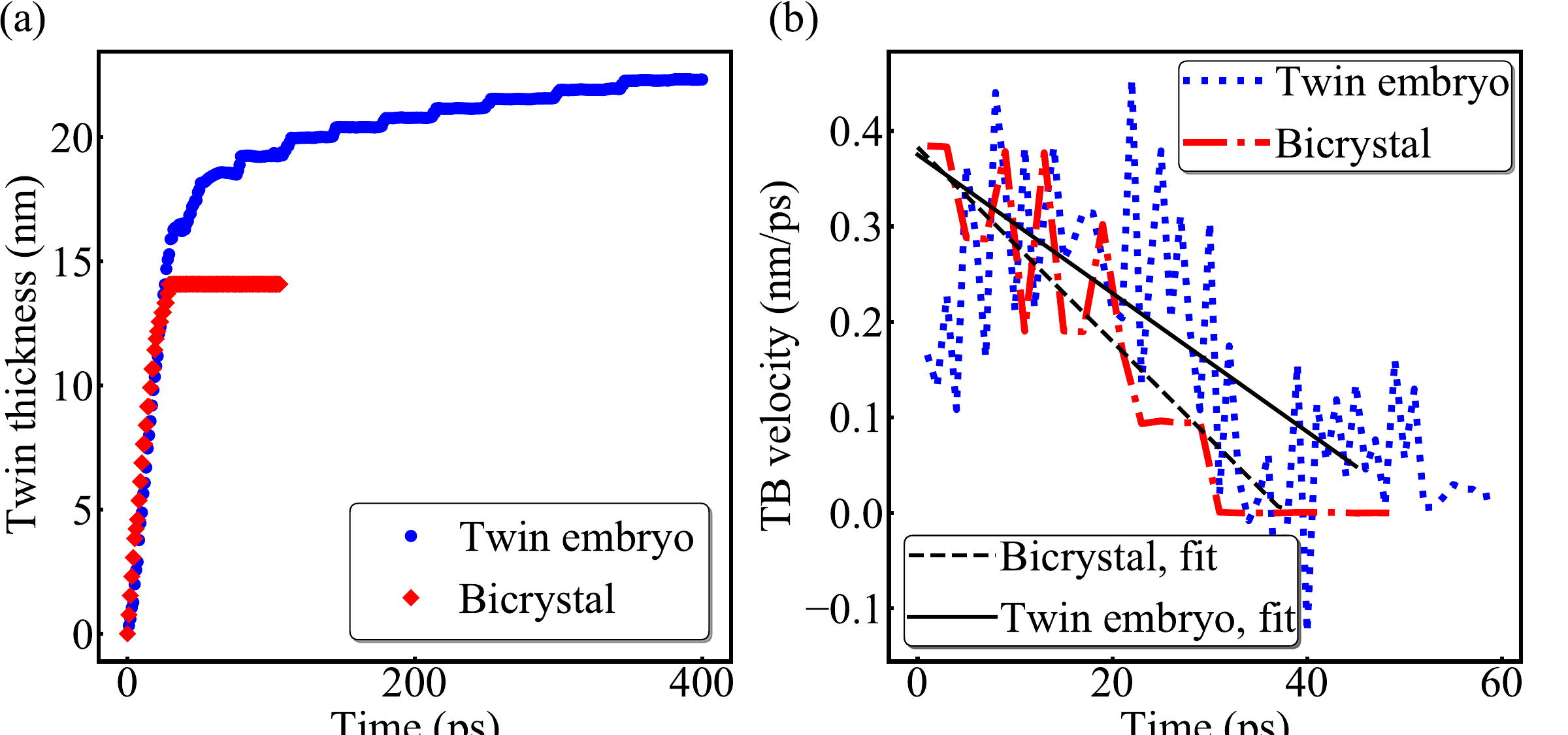}
  \caption{(a) Comparison of the time evolution of the twin thickness obtained by the twin embryo MD simulation and a bicrystal MD simulation with similar box height. For a fair comparison, the initial twin thickness (at 0 $ps$) is subtracted from the measured twin thickness. (b) Comparison of the time evolution of the TB velocity obtained by the twin embryo MD simulation and a bicrystal MD simulation with similar box height. Both instantaneous velocities and those estimated using quadratic regression are shown.}
  \label{Figure 6}
\end{figure}

Differences in facet motion are observed when they migrate as part of the boundary that separates the twin embryo from the parent versus when they migrate individually (\cref{Figure 6}).
For example, the motion of a f-TB is similar to the motion of a TB when they migrate individually, yet when a f-TB appears as part of the TT, its motion is faster.
When comparing the simulation of a twin embryo with the simulation of a bicrystal of the same box height ($\sim$77 $nm$) and the same applied shear strains, it is found that the motion of TBs is also faster in the twin embryo simulation and results in a larger final twin thickness.
This is evidence of BP/PB interfaces acting as a source of twinning disconnections.
Disconnections formed on the BP/PB interfaces have been found to transform into twinning disconnections on TBs and f-TBs \cite{hu2020disconnection,Barrett2014GB} and thus contribute to their motion.

When twin facets are connected (as in the case of a twin embryo), the stress fields caused by different facets interact with each other. Therefore, one may question whether or not the previously determined interface mobilities using bicrystals still apply to the motion of facets as part of the twin boundaries.
For an estimation of the TB and TT mobilities, slabs containing two TBs or two f-TBs are taken for the calculation of driving forces, and facet velocities are estimated using the quadratic regression of the time evolution of twin dimensions to calculate the interface mobilities.
Since facets of different types are too close to each other in small twin embryos, the data are collected after 20~ps, when the separation of TBs is around 1/3 of the simulation box height.
The obtained mobilities of f-TBs (as part of the TT) and TBs are about 4 times higher than those obtained using bicrystals. Yet the mobility ratio of TT to TB obtained by twin embryo simulation is about 4.5, and this value is comparable to the mobility ratio of BP/PB to TB (4-5, see \cref{Figure 3}) calculated using bicrystals. 

\section{Phase field modeling} \label{sec:phase_field_modeling}
This section presents the phase field disconnection model applied to the disconnection-mediated twin growth at the larger length and time scale, with the incorporation of interface energies of TBs and BP/PB interfaces as well as the nonlinear kinetic behavior of TBs and BP/PB interfaces obtained from MD simulations.
This extends the model originally reported in \cite{Chesser2020,Gokuli2021}, which are general references for this section. 

\subsection{Disconnection-mediated twin growth model}
The phase field disconnections model is a generalization of the principle of minimum dissipation potential (which is a variant of maximum dissipation) for GBs, which has been highly effective in modeling various non-equilibrium mechanical processes such as viscoplasticity \cite{Ortiz1999,Hackl2008,Carstensen2002,Tomas2010,Roubicek2009} and has been shown to capture aggregate GB migration in microstructure \cite{Chesser2020,bugas2024grain}.
The minimum dissipation potential framework, when written in the context of a phase field, may be stated as
\begin{equation}\label{eq:pfdd_master}
\inf_{\dot\eta \in \boldsymbol{C}^{4}(\Omega)^{N}} \left\{\frac{\partial}{\partial t}\Big(\inf_{\bm{u} \in\text{adm.}} \Pi[\boldsymbol{u}, \eta]\Big) + \Phi^{*} (\dot{\eta})\right\},
\end{equation}
where $\Pi$ is the total potential energy, $\Phi^{*}$ is the total dissipation, $\eta$ is a set of order parameters, and $\dot{\eta}=\frac{\partial \eta}{\partial t}$. $\boldsymbol{u}$ is the elastic displacement field, and dots denote time derivatives. $N$ is the number of grains or domains.
The use of brackets to indicate arguments is understood here to imply functional dependence upon the argument and any necessary spatial or temporal derivatives.
The restriction $\bm{u}\in\operatorname{adm.}$ enforces boundary conditions and continuity restrictions on the displacement field.
The continuity restriction $\eta\in C^4$ on the simulation domain $\Omega$ is necessary in the presence of strongly anisotropic interfaces (discussed subsequently).
The remainder of the section is a brief exposition of the different components of \cref{eq:pfdd_master}, which is the master equation of the phase field disconnection theory. In this work, only two order parameters are required to differentiate between twin and parent.

The free energy $W$ is decomposed into
\begin{equation}
W = W_{M}(\eta) + W_{B}(\nabla \eta) + W_{C}(\nabla^{2} \eta) + W_{E}(\eta, \nabla \bm{u}).
\end{equation}
Components $W_M$, $W_B$, $W_C$, and $W_E$ are, respectively, the chemical potential, the boundary energy, corner regularization, and elastic energy contributions, as described in the following.
The use of $\nabla$ in this work indicates the gradient, $\nabla^2$ the Hessian, and $\Delta=\operatorname{tr}(\nabla^2)$ the Laplacian.

The chemical multi-well potential $W_M$ is the same as in \cite{Moelans2008GB,Moelans2008} and 
is minimized when exactly one component of the order parameter is unity and the others are zero.
$W_{M}$ can be interpreted either as a mixing energy or as a segregation term.
The boundary energy $W_B$ represents the free energy contribution of the diffuse interface and has the form
\begin{equation}
W_{B}(\nabla \eta) = \frac{1}{2} \sum_{n=1}^{N} k(\boldsymbol{n}_{n}) |\nabla \eta_{n}|^{2},\qquad \boldsymbol{n}_{n} = \frac{\nabla \eta_{n}}{|\nabla \eta_{n}|},
\end{equation}
where $\boldsymbol{n}_{n}$ is the interface normal to domain $n$ along the boundary.
The coefficient $k(\boldsymbol{n}_{n})$ for the multiple-phase case \cite{Gokuli2021} is related to
the diffuse interface width $t_\text{int}$ and the orientation-dependent, strongly nonconvex interface energy via  $k(\boldsymbol{n}) = \frac{3t_\text{int}}{4}\sigma(\boldsymbol{n})$.
The calculation of $\sigma(\bm{n})$ is described below (\cref{sec:interface_energy_mobility_threshold}).

The strong nonconvexity of $\sigma$ is known for its role in creating microfacets on boundaries \cite{Sutton1995}, but also for inducing numerical challenges due to a lack of inherent length scale.
To overcome this issue, an additional term for curvature penalization is included, denoted as $W_{C}$.
$W_{C}$ depends on the second and third principle curvatures of $\eta$ \cite{Gokuli2021}, calculated from $\nabla^{2} \eta$. This particular form (originally derived in \cite{Ribot2019}) is beneficial, as it relies solely on the physical curvature rather than the diffuse curvature of the boundary. 

Finally, the elastic energy contribution $W_E$ is given by the quartic mixture rule
\begin{equation}
W_{E}(\eta,\nabla \bm{u}) = \frac{2\sum_{n=1}^{N} U_{n}(\nabla \bm{u}) \eta_{n}^{4}}{\sum_{n, m > n}^{N} \eta_{n}^{2}\eta_{m}^{2}},
\end{equation}
where $U_n$ is the elastic free energy contribution from domain $n$ dependent on the displacement gradient $\nabla\bm{u}$.
In this work, linearized elasticity about twin interface shear deformations is considered; this accounts for the large deformations induced by shear coupling without the usual expense of large-deformation mechanics.
The elastic free energies hence are
\begin{equation}
U_{n}(\nabla \boldsymbol{u}) = \frac{1}{2}(\nabla \boldsymbol{u}-\Delta\mathbf{F}_{n}^{GB}):\mathbb{C}(\nabla \boldsymbol{u}-\Delta\mathbf{F}_{n}^{GB}),
\end{equation}
where the deformation gradient tensor $\Delta\mathbf{F}_n^{GB}$ contains the shear associated with the shear-coupled motion of the boundary, i.e., the shear coupling factor.
In this study, the shear coupling factor is the twinning shear of the $\hkl{10-12}$ twin.
The elastic modulus tensor $\mathbb{C}$ for a rotated hexagonal material was used (as described in \cref{sec:calculating_driving_force}).

In addition to the free energy, the model is defined by the dual dissipation potential, which encapsulates the rate of energy dissipation as a function of the rate of the order parameter.
Here, the second-order form 
\begin{equation}\label{eq:DissPot}
\phi^{*}(\dot{\eta}) = \phi_{0}|\dot{\eta}_{n}|+\frac{1}{2}\phi_{1}\dot{\eta}_{n}^{2}
\end{equation}
is used, which is sufficient to model rate-dependent motion (governed by the inverse mobility $\phi_1$) with a critical driving force (determined by the constant $\phi_0$).
The inverse mobility coefficient is related to the other phase field parameters via
\begin{equation}
\phi_0=\left(\frac{4}{3} \frac{M}{t_\text{int}}\right)^{-1},
\end{equation}
where $M$ is the interface mobility (velocity over driving force), and $t_\text{int}$ is the diffuse interface width \cite{Moelans2008}.
Whereas previous applications of this model (targeting GB motion) assumed isotropic mobility, it is necessary to account for orientation dependence when considering the distinct kinetic behaviors associated with twin lengthening and thickening.

The governing equations for $\eta$ and $\boldsymbol{u}$ follow from \cref{eq:pfdd_master}.
The inner variational problem for the equilibrium displacement field $\bm{u}^*$ is solved first at every timestep.
Assuming quasistatic loading and no body forces, this amounts to solving the Euler-Lagrange equations
\begin{subequations}\begin{align}
\operatorname{Div}\Big(\frac{\partial W_{E}}{\partial \nabla \boldsymbol{u}}\Big) &= \boldsymbol{0} &&\forall\, \bm{X} \in \Omega \label{eq:stressdiv}\\
\boldsymbol{u} &= \boldsymbol{u}_{0} &&\forall \,\bm{X} \in \partial_{1} \Omega \label{eq:dispbc}\\
\frac{\partial W_{E}}{\partial \nabla \boldsymbol{u}} \boldsymbol{n} &= \boldsymbol{t}_{0} &&\forall \, \bm{X} \in \partial_{2} \Omega \label{eq:tracbc}
\end{align}\label{eq:elasticityeqs}\end{subequations}
which are the governing equations of elasticity subject to appropriate boundary conditions (prescribing displacements $\bm{u}_0$ and tractions $\bm{t}_0$ on boundaries $\partial\Omega_1,\partial\Omega_2\subset\partial\Omega$, respectively.
The methods for solving \cref{eq:stressdiv,eq:dispbc,eq:tracbc} computationally are discussed in \cref{sec:computation}. 

\begin{figure*}[!ht]
  \centering
  \includegraphics[width=0.9\textwidth]{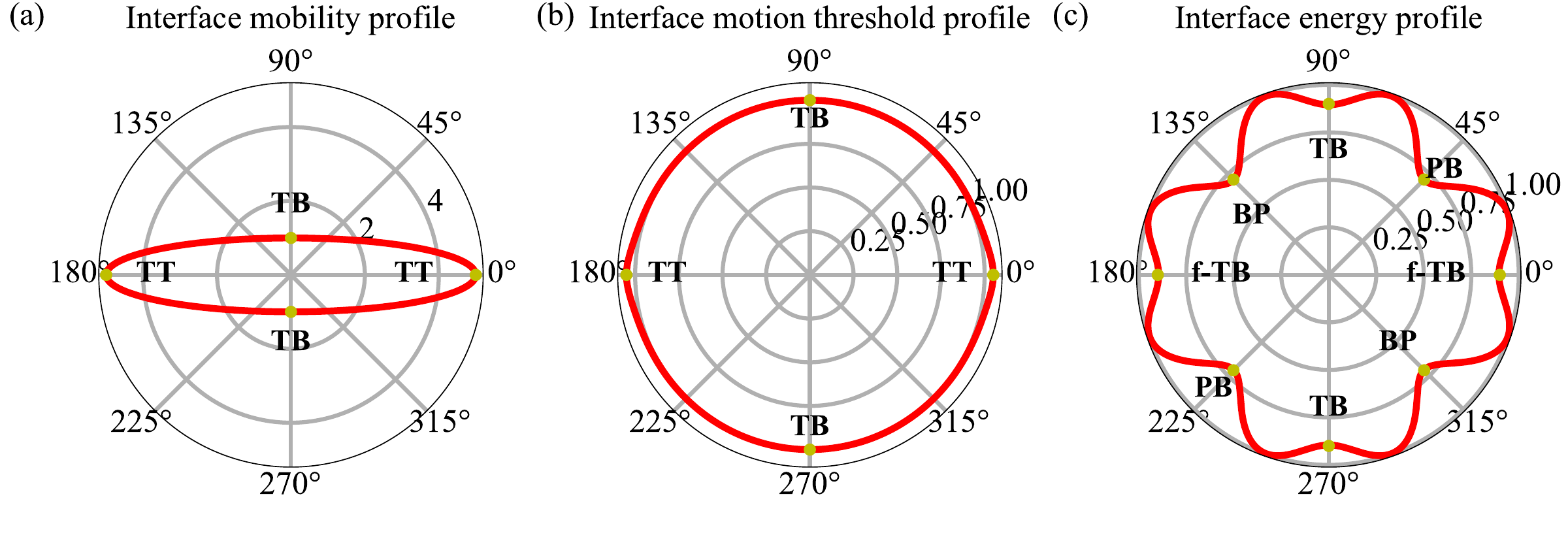}
  \caption{
    (a) Facet mobilities, (b) motion thresholds (eV/nm$^{3}$), and (c) interfacial energies (eV/nm$^{2}$) with respect to the interface orientation.
    Red lines correspond to interpolated data, yellow dots correspond to calibration points.
  }
  \label{Figure 7}
\end{figure*}

The order parameter $\eta$ is determined by the outer variational problem in \cref{eq:pfdd_master}, for which the Euler-Lagrange equation is
\begin{equation}
0 \in \frac{\delta}{\delta \eta_{n}} W(\boldsymbol{u}^{*}, \eta, \nabla \eta, \nabla^{2} \eta) + \frac{\partial}{\partial \dot{\eta}_{n}} \phi_{*}(\dot{\eta}),
\end{equation}
in which $\delta \slash \delta \eta$ is the variational derivative and $\partial \slash \partial \dot{\eta}$ is the subdifferential \cite{Rockafellar1997}.
Adopting the form in \cref{eq:DissPot} for $\phi^{*}$ allows the evolution equation to be simplified to
\begin{equation}
  \frac{\partial \eta_{n}}{\partial t} = -\frac{1}{\phi_{1}} 
    \begin{cases}
      \delta W \slash \delta \eta_{n}-\phi_{0} & \text{if } +\delta W \slash \delta \eta_{n} > +\phi_{0}, \\
      \delta W \slash \delta \eta_{n}+\phi_{0} & \text{if } -\delta W \slash \delta \eta_{n} < -\phi_{0}, \\
      0 & \text{else}.
    \end{cases}
\end{equation}
This amounts to a nonlinear kinetic relation for $\eta_n$ when $\phi_0>0$, and it recovers the usual Ginzburg-Landau-type $L_2$-gradient descent when $\phi_0=0$.
Nonlinear kinetics in phase field models are increasingly used to model complex behavior in diffuse interface systems \cite{Guin2023}.
In this particular case, the use of a threshold enables the existence of metastable states, which are not possible in usual phase field methods.
In the present work, both inverse mobility and dissipation energy are allowed to vary with interface normal, as discussed in the following section.

\subsection{Inclination-dependent properties}\label{sec:interface_energy_mobility_threshold}

At small scales, interfaces such as GBs and the twin boundaries studied here are understood to exhibit dependence on the orientation relationship of the boundary (sometimes called \textquotedblleft anisotropy\textquotedblright) as well as the boundary plane inclination itself (sometimes called \textquotedblleft orientation dependence\textquotedblright) \cite{Upmanyu1999,Gottstein1992,Molodov2022}.
The interface energy, in particular, usually exhibits many sharp minima (``cusps''), which are highly localized in both the space of misorientation and interface inclination.
It is also well-known that the interface mobility -- and, we posit, dissipation energy -- depends on the facet character \cite{Molodov2022,Chesser2020}.
In the present work, the only orientation relationship considered is that of the twin and parent.
Therefore, the complete set of energies, mobilities, and dissipation energies is determined as a function over the set of interface plane normal vectors $\bm{n}$.

The determination of inclination-dependent interface properties for any material is challenging; for HCP metals especially there is a lack of available atomistic data for property determination.
In this work, the interfacial energies of TB, f-TB, and BP/PB interfaces are calculated using MD simulations.
That information is used to estimate the entire range of mobility, threshold, and interface energy (\cref{Figure 7}).

Certain approximations are made in the determination of the inclination-dependent parameters.
For the mobility, it is assumed that there is a smooth variation from the f-TB to TB inclinations; even though, in reality, the relation is likely more complex.
For the threshold profile, calculations indicate almost no dependence on orientation, and it is assumed that this holds for all facet orientations.
Interestingly, this is consistent with the definition of the threshold as a critical driving force, as there is no notion of a plane or orientation in its definition.
For the interface energy, a larger number of points are sampled, yielding an ostensibly more reliable estimate.
However, it must be remarked that the cusps in the profile, which are normally sharp, are approximated as smooth.
This greatly improves the stability of the phase field method and does not substantially alter the thermodynamics of the boundary.

\subsection{Computation}\label{sec:computation}

All phase field methods require a sufficiently high grid resolution to accurately capture diffuse boundaries without mesh dependency.
The use of a coarse grid can be used phenomenologically to create the  appearance of faceting, but this approach does not accurately reflect the physics of the system.
To ensure that faceting is truly driven by continuum thermodynamics, it is essential to adequately resolve the diffuse boundary (to the point where additional resolution does not change the result).
Typical best practice mandates at least four to eight grid spacings across the diffuse boundary \cite{Sun2007}.
Such resolution in real-space methods incurs excessive computational expense, unless an adaptive meshing refinement (AMR) scheme is used.

Simulations in this work use block-structured adaptive mesh refinement (BSAMR) \cite{Deiterding2011}.
In contrast to other AMR strategies, such as decimation or quad/octree division, BSAMR uses a patch-based scheme to evolve refinement levels independently.
This reduces the communication overhead and enables efficient load balancing for exceptionally good scaling \cite{Zhang2019}.
This scheme also allows for temporal subcycling, allowing finer levels to undergo multiple iterations for each iteration on the coarser level.
This is helpful for ensuring stability in explicit time integration, especially with the presence of a fourth-order spatial derivative in the second-order regularization.
Regridding occurs every 10 timesteps (on the coarse grid), and the criterion for regridding is
\begin{equation}
  |\nabla \eta_{n}|\Delta x > r
\end{equation}
where $r$ is the refinement threshold (value listed in \cref{Table 3}) and $\Delta x$ is the AMR grid size.

For solving the quasistatic equilibrium equations, a strong form solver is used, which employs the \textquotedblleft reflux-free\textquotedblright{} method to treat the coarse-fine boundary.
This approach is particularly advantageous, as it allows for the use of the highly efficient geometric multigrid method, resulting in a very efficient solver \cite{Runnels2020,Runnels2021,Agrawal2023}.
The implicit elasticity and explicit phase field problems are solved using a staggered approach.
It is determined that the elastic solutions can conservatively be updated only every 10 timesteps without any appreciable impact on the result.
The phase field model is implemented in the C++ code Alamo \cite{alamo}, which uses the AMReX library to provide data structures and algorithms for BSAMR \cite{amrex}.

The periodic simulation box in phase field simulations is set up consistently with the MD simulations in \cref{sec:md_simulations}.
A twin domain of elliptical shape with width $\sim$9.0~nm and height $\sim$4.2~nm is centered in the simulation box.
To activate twin growth in phase field simulations, simulation boxes are deformed by applying average shear displacements along the $X$-axis at time $t = 0$.
The applied displacements are slightly adjusted for different simulations so that the initial twin volume fraction matches that of MD simulations.
All values of the phase field parameters are summarized in \cref{Table 4}. 

\begin{table}
  \centering
  \caption{AMR parameters}
  \label{Table 3}
  \begin{tabularx}{\linewidth}{XX}
    \toprule
    Parameter & Value \\
    \midrule
    Base grid	& 64$\times$64 \\
    Base timestep	& 0.0001 (equivalent to 1 fs) \\
    \# AMR levels & 4 \\ 
    Refinement threshold & $r$ = 0.1 \\
    \bottomrule
  \end{tabularx}
\end{table}

\begin{table}
  \centering
  \caption{Phase field parameters}
  \label{Table 4}
  \begin{tabularx}{\linewidth}{XX}
    \toprule
    Parameter & Value(s) \\
    \midrule
    Boundary width & $t_\text{int}$ = 0.5, 0.8, 1~nm \\
    Regularization energy & $\beta$ = 0.005, 0.01, 0.02~eV\\ 
    Top displacement & 3.4, 3.5~nm \\ 
    \bottomrule
  \end{tabularx}
\end{table}

\begin{figure*}
\centering
\includegraphics[width=0.9\textwidth]{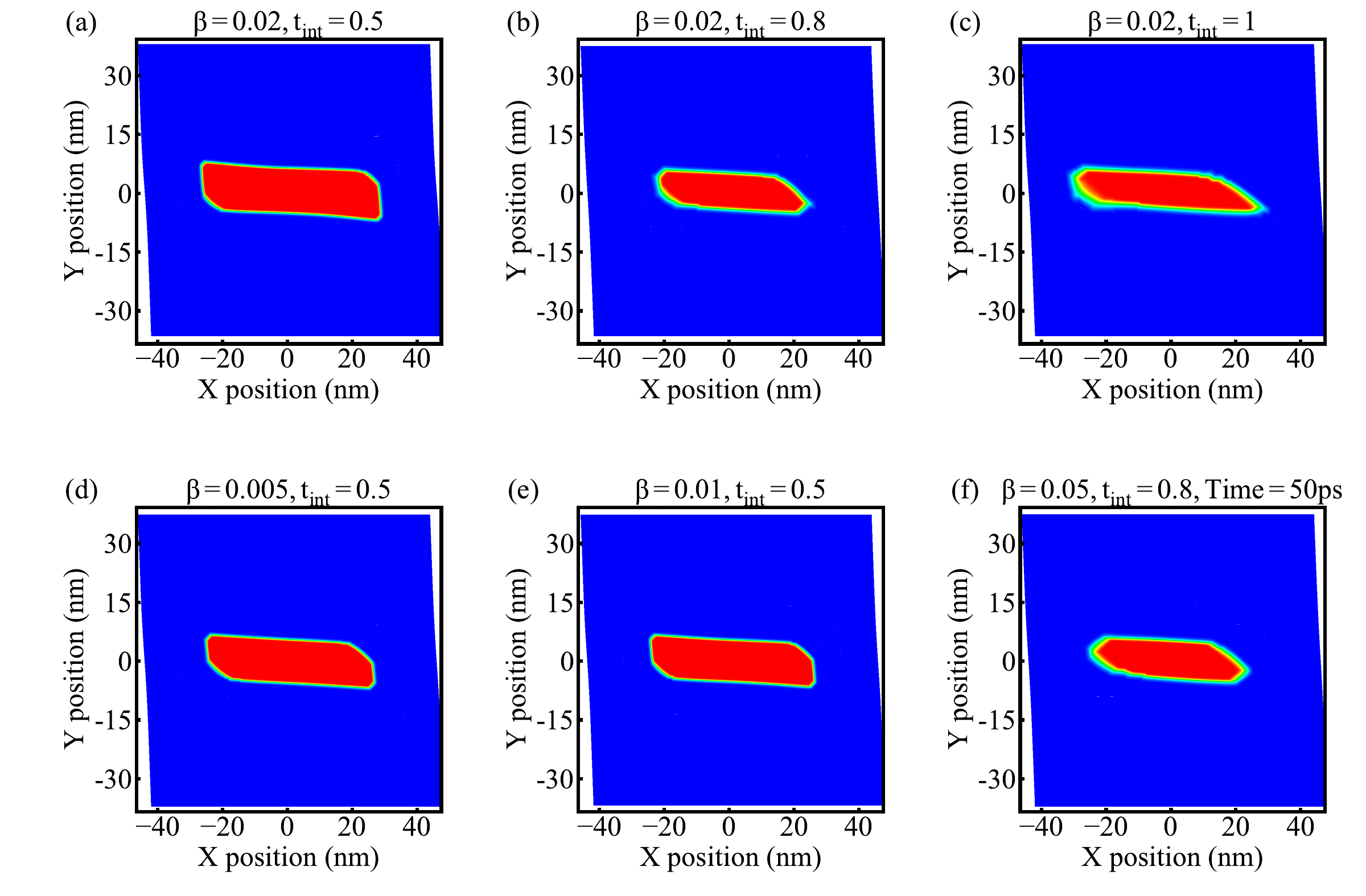}
\caption{The twin structure at 40~ps from phase field simulation with (a) $\beta=0.02$ and $t_{int}=0.5$, (b) $\beta=0.02$ and $t_{int}=0.8$, (c) $\beta=0.02$ and $t_{int}=1$, (d) $\beta=0.005$ and $t_{int}=0.5$, (e) $\beta=0.01$ and $t_{int}=0.5$, and (f) the twin structure at 50~ps from simulation with $\beta$=0.05 and $t_{int}$=0.8.}
\label{fig:pf_diff_parameter}
\end{figure*}

\subsection{Phase field results}

Most of the model parameters in this work are informed directly by MD data, except for two key values: the diffuse thickness $t_\text{int}$ and the regularization parameter $\beta$.
The diffuse interface thickness is a numerical regularization parameter and should in principle be sufficiently small to approach the sharp interface limit.
At the same time, best practice for diffuse interface methods \cite{schmidt2022self} indicates that the diffuse width should be at least 5-8 voxels (on the finest AMR level) to adequately resolve the diffuse boundary, which places computational limits on the minimum feasible thickness.
In practice, $t_\text{int}$ acts as a lower bound on resolvable length scales; only features with characteristic lengths larger than $t_\text{int}$ can be captured.
Practically, therefore, it is left as a variable parameter, whose effect on the overall result is determined in an Edisonian fashion.
The other undetermined parameter is the regularization energy $\beta$, which corresponds approximately to the core energy of a disconnection or facet.
Because of the inherent difficulty in obtaining a precise direct measurement for $\beta$ from atomistic simulations, we again leave it to be determined by comparison between PF and MD results.

The resulting phase field simulations are performed for a selection of parameters $\beta$ and $t_\text{int}$ each of which leads to slightly differing twin microstructure (\cref{fig:pf_diff_parameter}).
All simulations clearly capture the existence of BP/PB interfaces, TBs, and f-TBs.
However, differences in simulation results are also observed, showing that $\beta$ and $t_\text{int}$ are closely associated with the growth behavior of the twin.
When the same $\beta$ but different $t_\text{int}$ are used, as the $t_\text{int}$ increases, a sharper TT (connection of one BP and one PB interface and no f-TB appears at the TT) and more disconnections on the interfaces are observed. 
When the same $t_\text{int}$ but different $\beta$ is used, the larger the $\beta$ value, the slower the twin growth , and no significant change in the twin microstructure is observed. 
The structure of the TT, either one f-TB connected with one BP and one PB interface, or the direct connection of one BP with one PB interface is consistent with experimental observations \cite{Tu2013,Tu2015,Tu2016}.  
It is concluded that values of $\beta=-0.02$ and $t_\text{int}$ produce a sufficiently close morphology without incurring excessive computational cost.


\section{Discussion} \label{sec:discussion}
\begin{figure*}
\centering
\includegraphics[width=0.9\textwidth]{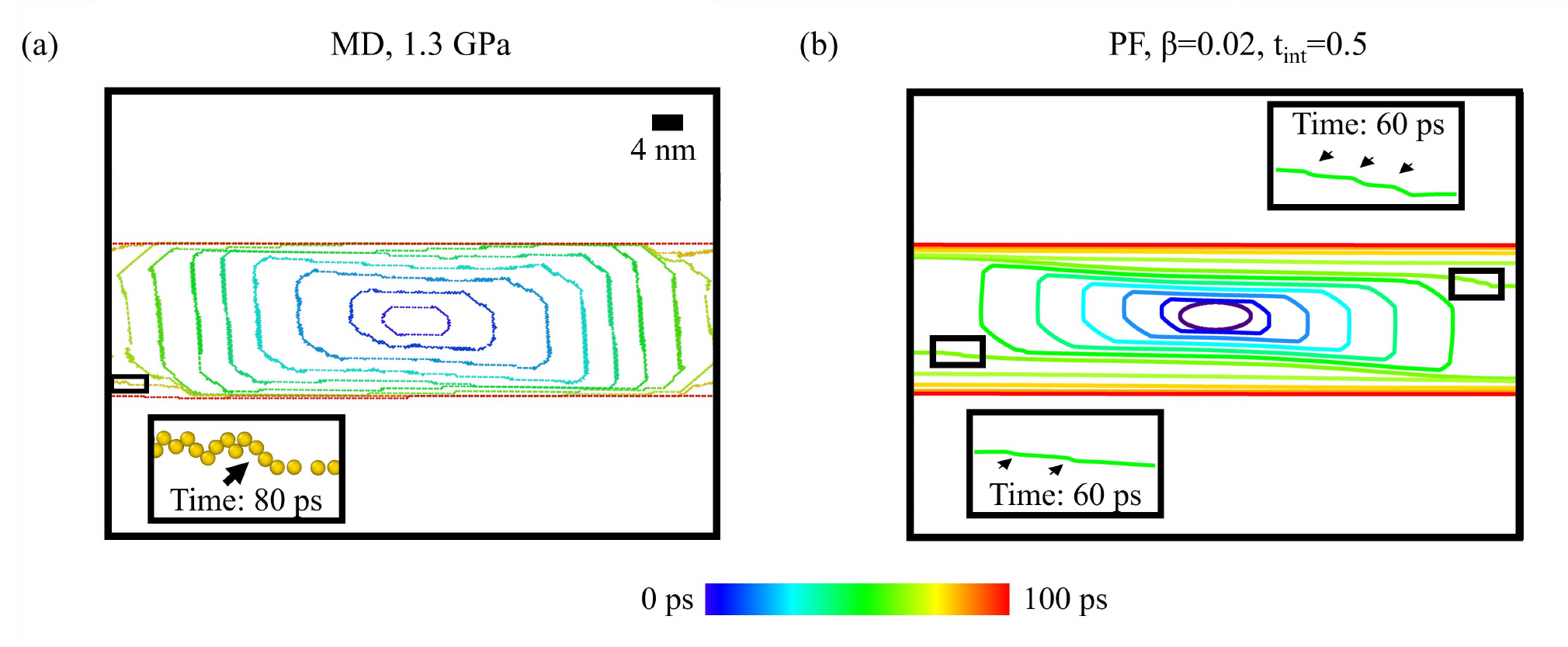}
\caption{Time evolution of the boundary that separates the twin from the parent colored by corresponding simulation time extracted from (a) MD simulations and (b) phase field simulations, using $\beta=0.02$ and $t_{int}=0.5$. Zoomed-in views of disconnections formed on the TBs and BP/PB interfaces at 10~ps, 30~ps, and 60~ps in MD simulations are shown in (a), and those formed at 60 ps in phase field simulations are shown in (b).}
\label{fig:contours_pf_and_md}
\end{figure*}

Comparison of the PF-predicted twin growth with the MD results shows good qualitative agreement (\cref{fig:contours_pf_and_md}).
The lengthening of the twin along the $X$-axis is, correctly, faster than the twin thickening along the $Y$-axis.
Furthermore, the overall twin shape exhibits the same facets with similar proportions. 
There is a slight temporal offset between the PF and MD results between 10-20 ps; this is attributed to the high sensitivity of the early growth rate on initial conditions.

Of particular note in this comparison is the emergence of distinct steps on twin facets (\cref{fig:contours_pf_and_md} popouts), which match closely with the disconnections observed in MD simulations.
In the PF simulations, the disconnections appear near the BP/PB interfaces; in MD, the majority of disconnections are formed at the intersections of TBs or f-TBs with BP/PB interfaces.
Nevertheless, disconnections formed at locations far from BP/PB interfaces are also observed in MD simulations.

\begin{figure*}
\centering
\includegraphics[width=0.9\textwidth]{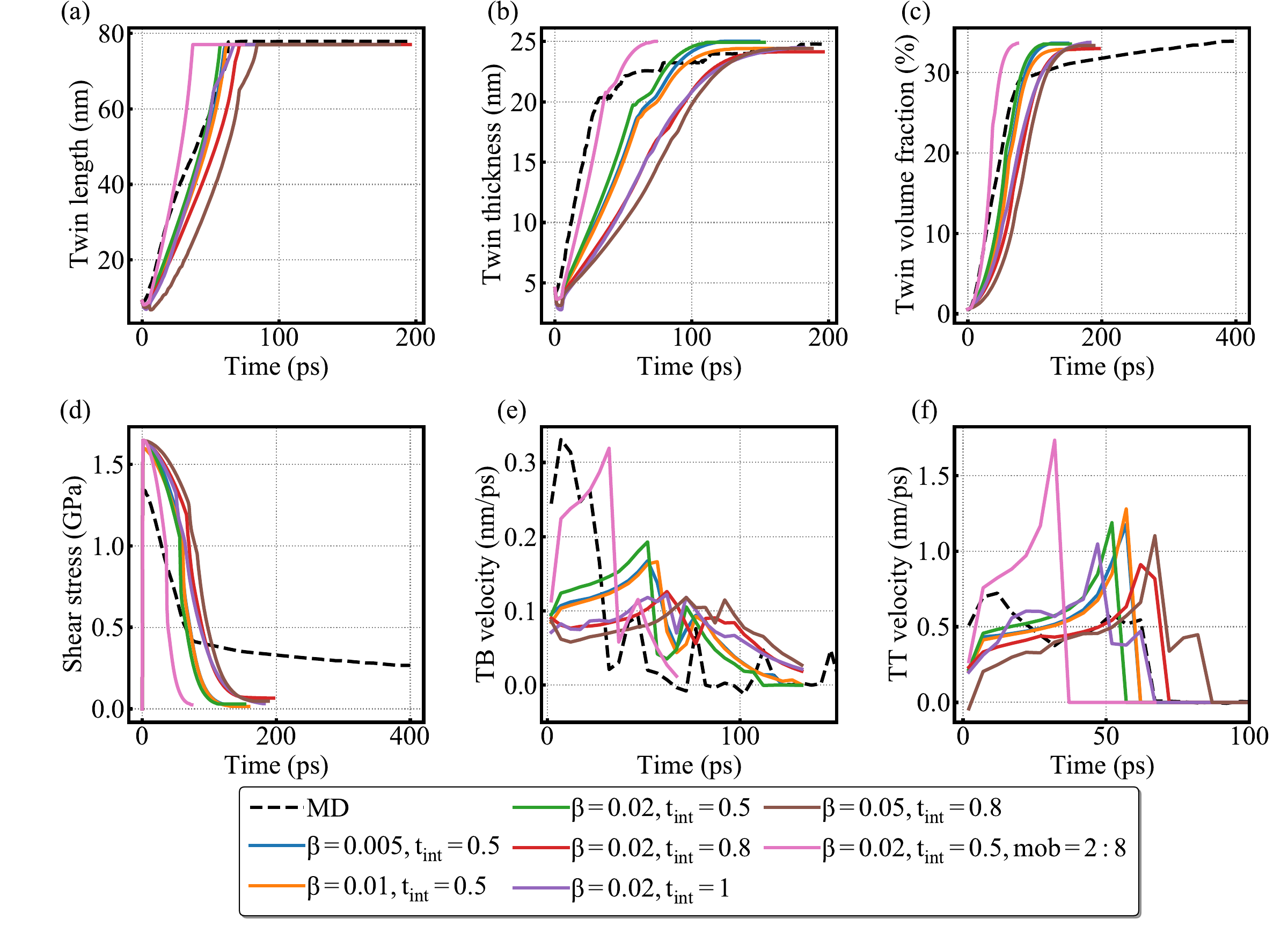}
\caption{Time evolution of the (a) twin length, (b) twin thickness, (c) twin volume fraction, (d) global shear stress, (e) TB velocity, and (f) TT velocity, compared with MD results.}
\label{Figure 10}
\end{figure*}

The time evolution of twin lengths and thicknesses, twin volume fraction, global shear stresses, and twin facet velocities are also extracted from phase field simulations with different parameters and compared with MD results (\cref{Figure 10}a-c). 
As with the analysis of the MD results, twin length and thickness are approximated by the differences between the maximum and minimum positions of the twin along the $X$- and $Y$-axes, respectively. 
Facet velocities are calculated using linear fits of every five data points from the curves of twin dimensions versus time. 
Reasonable agreement between MD and phase field simulations is shown in the final twin length and thickness as well as the time evolution of the twin volume fraction. 
Within the initial 2-3~ps of phase field simulations, there is a slight decrease in the measured twin length and thickness due to the change of twin shape from the initial elliptical shape, and there is no actual decrease in the twin size. 
Overall, there is a reasonable quantitative match between the MD and PF results for the geometry of the embryo.

The mechanical response of the PF and MD systems are compared by plotting the relaxation of the shear stress (averaged over the top face) vs time (\cref{Figure 10}d).
Both feature a rapid decrease of the global shear stress  within $\sim$65~ps.
The stress in the PF model nearly completely relaxes.
The MD stress, on the other hand, levels out considerably, leaving much more substantial residual shear stress.
This is due to the more limited effect of periodic images on the driving force in MD, as well as to the presence of additional relaxation mechanisms (such as plastic slip or interface sliding) that is inaccessible to the PF.
Nevertheless, given enough time, the global shear stresses in MD simulations eventually do vanish. 

The MD and PF facet velocities are also compared (\cref{Figure 10}e-f).
Both show a significant decrease of both TT and TB velocities after $\sim$65~ps, and the small peaks appearing in the curve for TB velocity again show the step-by-step increase of twin thickness after $\sim$65~ps. 
The phase field simulations with a TB mobility of 2 and TT mobility of 8 (where mobility is measured as interface velocity divided by interface driving force, expressed in the unit of $\text{nm}^{4}/(\text{eV}\cdot \text{ps})$) yield an overall time evolution of TB and TT velocities which is quantitatively consistent with MD simulations. 
These mobility values are critical as they govern the rate at which twins propagate through the grain, ensuring that the phase field simulations mirror the dynamic behaviors captured in the atomistic simulations.

In general, both qualitative agreement is observed through the overall shape and growth rate of the twin nucleus in MD and PF, and there is quantitative agreement through the comparison of stresses, facet positions, and facet velocities.
Except for the corner regularization and diffuse thickness parameters, all parameters were fit using separate bicrystal simulations.

\section{Conclusion} \label{sec:conclusion}

This work demonstrates the ability of the fully resolved phase field disconnection modeling approach, with nonlinear kinetics and strong GB anisotropy, to capture the behavior of Mg twin growth mediated by disconnections. 
While twin growth in mg has been reported in previous works (e.g. \cite{spearot2020structure,gong2021effects}), this work employs the essential combined mechanisms of nonlinear kinetics, strongly anisotropic boundary properties, corner regularization, and adaptive mesh refinement to ensure that the solution is free of numerical or mesh-related artifacts.
In addition to its numerical robustness, this phase field model is uniquely capable of capturing facet behavior as an ``emergent'' phenomenon, as can be seen by the apparent disconnection activity in the PF model on the twin boundary.

The accuracy of the model is underscored by the MD simulations that are used to inform the model parameters.
The MD simulations reported here indicate exactly the type of nonlinear kinetic response behavior suggested in \cite{Runnels2020,Chesser2020,Gokuli2021}.
In addition to GB energies and critical driving forces, anisotropic mobilities of BP/PB interfaces, calculated from bicrystal MD simulations, can be attributed to the observed rate difference between twin lengthening than twin thickening.

Despite the good agreement between the PF and MD results, a number of limitations in the model are acknowledged.
First, it is noted that the results presented here are 2D.
Though there are no conceptual limitations to performing a 3D study, the extensive atomistic simulations required to fully capture 3D anisotropy for all parameters would constitute an entirely separate study. 
Given the close match between phase field and MD simulations in 2D results and their potential implications, the efficacy of the model in 2D is convincingly demonstrated. 
Importantly, the relative ease and lower computational cost of the phase field model compared to 3D MD simulations suggest that this model could serve as a valuable and convenient approach for exploring 3D phenomena in future studies.

\section{Acknowledgments}
The support from the European Research Council (ERC) under the European Union's Horizon 2020 research and innovation program (grant agreement no. 770754) is gratefully acknowledged.
BR acknowledges support from the United States National Science Foundation, grant number MOMS-2341922.
This work used the INCLINE cluster at the University of Colorado Colorado Springs. INCLINE is supported by the National Science Foundation, grant number OAC-2017917.

\bibliography{main.bib}

\begin{thebibliography}{79}%
\makeatletter
\providecommand \@ifxundefined [1]{%
 \@ifx{#1\undefined}
}%
\providecommand \@ifnum [1]{%
 \ifnum #1\expandafter \@firstoftwo
 \else \expandafter \@secondoftwo
 \fi
}%
\providecommand \@ifx [1]{%
 \ifx #1\expandafter \@firstoftwo
 \else \expandafter \@secondoftwo
 \fi
}%
\providecommand \natexlab [1]{#1}%
\providecommand \enquote  [1]{``#1''}%
\providecommand \bibnamefont  [1]{#1}%
\providecommand \bibfnamefont [1]{#1}%
\providecommand \citenamefont [1]{#1}%
\providecommand \href@noop [0]{\@secondoftwo}%
\providecommand \href [0]{\begingroup \@sanitize@url \@href}%
\providecommand \@href[1]{\@@startlink{#1}\@@href}%
\providecommand \@@href[1]{\endgroup#1\@@endlink}%
\providecommand \@sanitize@url [0]{\catcode `\\12\catcode `\$12\catcode
  `\&12\catcode `\#12\catcode `\^12\catcode `\_12\catcode `\%12\relax}%
\providecommand \@@startlink[1]{}%
\providecommand \@@endlink[0]{}%
\providecommand \url  [0]{\begingroup\@sanitize@url \@url }%
\providecommand \@url [1]{\endgroup\@href {#1}{\urlprefix }}%
\providecommand \urlprefix  [0]{URL }%
\providecommand \Eprint [0]{\href }%
\providecommand \doibase [0]{https://doi.org/}%
\providecommand \selectlanguage [0]{\@gobble}%
\providecommand \bibinfo  [0]{\@secondoftwo}%
\providecommand \bibfield  [0]{\@secondoftwo}%
\providecommand \translation [1]{[#1]}%
\providecommand \BibitemOpen [0]{}%
\providecommand \bibitemStop [0]{}%
\providecommand \bibitemNoStop [0]{.\EOS\space}%
\providecommand \EOS [0]{\spacefactor3000\relax}%
\providecommand \BibitemShut  [1]{\csname bibitem#1\endcsname}%
\let\auto@bib@innerbib\@empty
\bibitem [{\citenamefont {Koike}(2005)}]{Koike2005}%
  \BibitemOpen
  \bibfield  {author} {\bibinfo {author} {\bibfnamefont {J.}~\bibnamefont
  {Koike}},\ }\bibfield  {title} {\bibinfo {title} {Enhanced deformation
  mechanisms by anisotropic plasticity in polycrystalline mg alloys at room
  temperature},\ }\bibfield  {journal} {\bibinfo  {journal} {Metallurgical and
  Materials Transactions A: Physical Metallurgy and Materials Science}\
  }\textbf {\bibinfo {volume} {36}},\ \href
  {https://doi.org/10.1007/s11661-005-0032-4} {10.1007/s11661-005-0032-4}
  (\bibinfo {year} {2005})\BibitemShut {NoStop}%
\bibitem [{\citenamefont {Reed-Hill}\ and\ \citenamefont
  {Robertson}(1957)}]{Reed-Hill1957}%
  \BibitemOpen
  \bibfield  {author} {\bibinfo {author} {\bibfnamefont {R.}~\bibnamefont
  {Reed-Hill}}\ and\ \bibinfo {author} {\bibfnamefont {W.}~\bibnamefont
  {Robertson}},\ }\bibfield  {title} {\bibinfo {title} {Additional modes of
  deformation twinning in magnesium},\ }\href
  {https://doi.org/10.1016/0001-6160(57)90074-3} {\bibfield  {journal}
  {\bibinfo  {journal} {Acta Metallurgica}\ }\textbf {\bibinfo {volume} {5}},\
  \bibinfo {pages} {717} (\bibinfo {year} {1957})}\BibitemShut {NoStop}%
\bibitem [{\citenamefont {Zhang}\ \emph {et~al.}(2021)\citenamefont {Zhang},
  \citenamefont {Li}, \citenamefont {Jiang}, \citenamefont {Li}, \citenamefont
  {Li}, \citenamefont {Zhang},\ and\ \citenamefont {Suo}}]{Zhang2021}%
  \BibitemOpen
  \bibfield  {author} {\bibinfo {author} {\bibfnamefont {Q.}~\bibnamefont
  {Zhang}}, \bibinfo {author} {\bibfnamefont {J.}~\bibnamefont {Li}}, \bibinfo
  {author} {\bibfnamefont {K.}~\bibnamefont {Jiang}}, \bibinfo {author}
  {\bibfnamefont {P.}~\bibnamefont {Li}}, \bibinfo {author} {\bibfnamefont
  {Y.}~\bibnamefont {Li}}, \bibinfo {author} {\bibfnamefont {Y.}~\bibnamefont
  {Zhang}},\ and\ \bibinfo {author} {\bibfnamefont {T.}~\bibnamefont {Suo}},\
  }\bibfield  {title} {\bibinfo {title} {Gradient structure induced
  simultaneous enhancement of strength and ductility in az31 mg alloy with
  twin-twin interactions},\ }\bibfield  {journal} {\bibinfo  {journal} {Journal
  of Magnesium and Alloys}\ }\href {https://doi.org/10.1016/j.jma.2021.10.014}
  {10.1016/j.jma.2021.10.014} (\bibinfo {year} {2021})\BibitemShut {NoStop}%
\bibitem [{\citenamefont {Wang}\ \emph
  {et~al.}(2020{\natexlab{a}})\citenamefont {Wang}, \citenamefont {Jiang},
  \citenamefont {Cooper}, \citenamefont {Yu}, \citenamefont {Zhang},
  \citenamefont {Rupert}, \citenamefont {Mahajan}, \citenamefont {Beyerlein},
  \citenamefont {Lavernia},\ and\ \citenamefont {Schoenung}}]{Wang2020}%
  \BibitemOpen
  \bibfield  {author} {\bibinfo {author} {\bibfnamefont {X.}~\bibnamefont
  {Wang}}, \bibinfo {author} {\bibfnamefont {L.}~\bibnamefont {Jiang}},
  \bibinfo {author} {\bibfnamefont {C.}~\bibnamefont {Cooper}}, \bibinfo
  {author} {\bibfnamefont {K.}~\bibnamefont {Yu}}, \bibinfo {author}
  {\bibfnamefont {D.}~\bibnamefont {Zhang}}, \bibinfo {author} {\bibfnamefont
  {T.~J.}\ \bibnamefont {Rupert}}, \bibinfo {author} {\bibfnamefont
  {S.}~\bibnamefont {Mahajan}}, \bibinfo {author} {\bibfnamefont {I.~J.}\
  \bibnamefont {Beyerlein}}, \bibinfo {author} {\bibfnamefont {E.~J.}\
  \bibnamefont {Lavernia}},\ and\ \bibinfo {author} {\bibfnamefont {J.~M.}\
  \bibnamefont {Schoenung}},\ }\bibfield  {title} {\bibinfo {title} {Toughening
  magnesium with gradient twin meshes},\ }\href
  {https://doi.org/10.1016/j.actamat.2020.05.021} {\bibfield  {journal}
  {\bibinfo  {journal} {Acta Materialia}\ }\textbf {\bibinfo {volume} {195}},\
  \bibinfo {pages} {468} (\bibinfo {year} {2020}{\natexlab{a}})}\BibitemShut
  {NoStop}%
\bibitem [{\citenamefont {Chen}\ \emph {et~al.}(2021)\citenamefont {Chen},
  \citenamefont {Ye}, \citenamefont {Wang}, \citenamefont {Zhou}, \citenamefont
  {Suo}, \citenamefont {Deng}, \citenamefont {Zhao},\ and\ \citenamefont
  {Wang}}]{Chen2021}%
  \BibitemOpen
  \bibfield  {author} {\bibinfo {author} {\bibfnamefont {L.}~\bibnamefont
  {Chen}}, \bibinfo {author} {\bibfnamefont {T.}~\bibnamefont {Ye}}, \bibinfo
  {author} {\bibfnamefont {Y.}~\bibnamefont {Wang}}, \bibinfo {author}
  {\bibfnamefont {D.}~\bibnamefont {Zhou}}, \bibinfo {author} {\bibfnamefont
  {T.}~\bibnamefont {Suo}}, \bibinfo {author} {\bibfnamefont {Q.}~\bibnamefont
  {Deng}}, \bibinfo {author} {\bibfnamefont {F.}~\bibnamefont {Zhao}},\ and\
  \bibinfo {author} {\bibfnamefont {Q.}~\bibnamefont {Wang}},\ }\bibfield
  {title} {\bibinfo {title} {Development of mechanical properties in az31
  magnesium alloy processed by cold dynamic extrusion},\ }\href
  {https://doi.org/10.1016/j.matchar.2021.111535} {\bibfield  {journal}
  {\bibinfo  {journal} {Materials Characterization}\ }\textbf {\bibinfo
  {volume} {182}},\ \bibinfo {pages} {111535} (\bibinfo {year}
  {2021})}\BibitemShut {NoStop}%
\bibitem [{\citenamefont {Tu}\ \emph {et~al.}(2013)\citenamefont {Tu},
  \citenamefont {Zhang}, \citenamefont {Wang}, \citenamefont {Sun},
  \citenamefont {Liu},\ and\ \citenamefont {Tomé}}]{Tu2013}%
  \BibitemOpen
  \bibfield  {author} {\bibinfo {author} {\bibfnamefont {J.}~\bibnamefont
  {Tu}}, \bibinfo {author} {\bibfnamefont {X.}~\bibnamefont {Zhang}}, \bibinfo
  {author} {\bibfnamefont {J.}~\bibnamefont {Wang}}, \bibinfo {author}
  {\bibfnamefont {Q.}~\bibnamefont {Sun}}, \bibinfo {author} {\bibfnamefont
  {Q.}~\bibnamefont {Liu}},\ and\ \bibinfo {author} {\bibfnamefont {C.~N.}\
  \bibnamefont {Tomé}},\ }\bibfield  {title} {\bibinfo {title} {Structural
  characterization of \{10-12\} twin boundaries in cobalt},\ }\bibfield
  {journal} {\bibinfo  {journal} {Applied Physics Letters}\ }\textbf {\bibinfo
  {volume} {103}},\ \href {https://doi.org/10.1063/1.4817180}
  {10.1063/1.4817180} (\bibinfo {year} {2013})\BibitemShut {NoStop}%
\bibitem [{\citenamefont {Tu}\ \emph {et~al.}(2015)\citenamefont {Tu},
  \citenamefont {Zhang}, \citenamefont {Ren}, \citenamefont {Sun},\ and\
  \citenamefont {Liu}}]{Tu2015}%
  \BibitemOpen
  \bibfield  {author} {\bibinfo {author} {\bibfnamefont {J.}~\bibnamefont
  {Tu}}, \bibinfo {author} {\bibfnamefont {X.~Y.}\ \bibnamefont {Zhang}},
  \bibinfo {author} {\bibfnamefont {Y.}~\bibnamefont {Ren}}, \bibinfo {author}
  {\bibfnamefont {Q.}~\bibnamefont {Sun}},\ and\ \bibinfo {author}
  {\bibfnamefont {Q.}~\bibnamefont {Liu}},\ }\bibfield  {title} {\bibinfo
  {title} {Structural characterization of \{10-12\} irregular-shaped twinning
  boundary in hexagonal close-packed metals},\ }\bibfield  {journal} {\bibinfo
  {journal} {Materials Characterization}\ }\textbf {\bibinfo {volume} {106}},\
  \href {https://doi.org/10.1016/j.matchar.2015.05.032}
  {10.1016/j.matchar.2015.05.032} (\bibinfo {year} {2015})\BibitemShut
  {NoStop}%
\bibitem [{\citenamefont {Tu}\ and\ \citenamefont {Zhang}(2016)}]{Tu2016}%
  \BibitemOpen
  \bibfield  {author} {\bibinfo {author} {\bibfnamefont {J.}~\bibnamefont
  {Tu}}\ and\ \bibinfo {author} {\bibfnamefont {S.}~\bibnamefont {Zhang}},\
  }\bibfield  {title} {\bibinfo {title} {On the \{10-12\} twinning growth
  mechanism in hexagonal close-packed metals},\ }\bibfield  {journal} {\bibinfo
   {journal} {Materials and Design}\ }\textbf {\bibinfo {volume} {96}},\ \href
  {https://doi.org/10.1016/j.matdes.2016.02.002} {10.1016/j.matdes.2016.02.002}
  (\bibinfo {year} {2016})\BibitemShut {NoStop}%
\bibitem [{\citenamefont {Liu}\ \emph {et~al.}(2015)\citenamefont {Liu},
  \citenamefont {Wan}, \citenamefont {Wang}, \citenamefont {Ma},\ and\
  \citenamefont {Shan}}]{Liu2015}%
  \BibitemOpen
  \bibfield  {author} {\bibinfo {author} {\bibfnamefont {B.~Y.}\ \bibnamefont
  {Liu}}, \bibinfo {author} {\bibfnamefont {L.}~\bibnamefont {Wan}}, \bibinfo
  {author} {\bibfnamefont {J.}~\bibnamefont {Wang}}, \bibinfo {author}
  {\bibfnamefont {E.}~\bibnamefont {Ma}},\ and\ \bibinfo {author}
  {\bibfnamefont {Z.~W.}\ \bibnamefont {Shan}},\ }\bibfield  {title} {\bibinfo
  {title} {Terrace-like morphology of the boundary created through
  basal-prismatic transformation in magnesium},\ }\bibfield  {journal}
  {\bibinfo  {journal} {Scripta Materialia}\ }\textbf {\bibinfo {volume}
  {100}},\ \href {https://doi.org/10.1016/j.scriptamat.2014.12.020}
  {10.1016/j.scriptamat.2014.12.020} (\bibinfo {year} {2015})\BibitemShut
  {NoStop}%
\bibitem [{\citenamefont {Wang}\ \emph
  {et~al.}(2020{\natexlab{b}})\citenamefont {Wang}, \citenamefont {Gong},
  \citenamefont {McCabe}, \citenamefont {Capolungo}, \citenamefont {Wang},\
  and\ \citenamefont {Tomé}}]{Wang2020characteristic}%
  \BibitemOpen
  \bibfield  {author} {\bibinfo {author} {\bibfnamefont {S.}~\bibnamefont
  {Wang}}, \bibinfo {author} {\bibfnamefont {M.}~\bibnamefont {Gong}}, \bibinfo
  {author} {\bibfnamefont {R.~J.}\ \bibnamefont {McCabe}}, \bibinfo {author}
  {\bibfnamefont {L.}~\bibnamefont {Capolungo}}, \bibinfo {author}
  {\bibfnamefont {J.}~\bibnamefont {Wang}},\ and\ \bibinfo {author}
  {\bibfnamefont {C.~N.}\ \bibnamefont {Tomé}},\ }\bibfield  {title} {\bibinfo
  {title} {Characteristic boundaries associated with three-dimensional twins in
  hexagonal metals},\ }\href {https://doi.org/10.1126/sciadv.aaz2600}
  {\bibfield  {journal} {\bibinfo  {journal} {Science advances}\ }\textbf
  {\bibinfo {volume} {6}},\ \bibinfo {pages} {eaaz2600} (\bibinfo {year}
  {2020}{\natexlab{b}})}\BibitemShut {NoStop}%
\bibitem [{\citenamefont {Wang}\ \emph {et~al.}(2022)\citenamefont {Wang},
  \citenamefont {Hu}, \citenamefont {Yu}, \citenamefont {Mahajan},
  \citenamefont {Beyerlein}, \citenamefont {Lavernia}, \citenamefont {Rupert},\
  and\ \citenamefont {Schoenung}}]{Wang2022}%
  \BibitemOpen
  \bibfield  {author} {\bibinfo {author} {\bibfnamefont {X.}~\bibnamefont
  {Wang}}, \bibinfo {author} {\bibfnamefont {Y.}~\bibnamefont {Hu}}, \bibinfo
  {author} {\bibfnamefont {K.}~\bibnamefont {Yu}}, \bibinfo {author}
  {\bibfnamefont {S.}~\bibnamefont {Mahajan}}, \bibinfo {author} {\bibfnamefont
  {I.~J.}\ \bibnamefont {Beyerlein}}, \bibinfo {author} {\bibfnamefont {E.~J.}\
  \bibnamefont {Lavernia}}, \bibinfo {author} {\bibfnamefont {T.~J.}\
  \bibnamefont {Rupert}},\ and\ \bibinfo {author} {\bibfnamefont {J.~M.}\
  \bibnamefont {Schoenung}},\ }\bibfield  {title} {\bibinfo {title} {Room
  temperature deformation-induced solute segregation and its impact on twin
  boundary mobility in a mg-y alloy},\ }\bibfield  {journal} {\bibinfo
  {journal} {Scripta Materialia}\ }\textbf {\bibinfo {volume} {209}},\ \href
  {https://doi.org/10.1016/j.scriptamat.2021.114375}
  {10.1016/j.scriptamat.2021.114375} (\bibinfo {year} {2022})\BibitemShut
  {NoStop}%
\bibitem [{\citenamefont {Zu}\ \emph {et~al.}(2017)\citenamefont {Zu},
  \citenamefont {Tang}, \citenamefont {Xu},\ and\ \citenamefont
  {Guo}}]{Zu2017}%
  \BibitemOpen
  \bibfield  {author} {\bibinfo {author} {\bibfnamefont {Q.}~\bibnamefont
  {Zu}}, \bibinfo {author} {\bibfnamefont {X.-Z.}\ \bibnamefont {Tang}},
  \bibinfo {author} {\bibfnamefont {S.}~\bibnamefont {Xu}},\ and\ \bibinfo
  {author} {\bibfnamefont {Y.-F.}\ \bibnamefont {Guo}},\ }\bibfield  {title}
  {\bibinfo {title} {Atomistic study of nucleation and migration of the
  basal/prismatic interfaces in mg single crystals},\ }\href
  {https://doi.org/10.1016/j.actamat.2017.03.035} {\bibfield  {journal}
  {\bibinfo  {journal} {Acta Materialia}\ }\textbf {\bibinfo {volume} {130}},\
  \bibinfo {pages} {310} (\bibinfo {year} {2017})}\BibitemShut {NoStop}%
\bibitem [{\citenamefont {Sun}\ \emph {et~al.}(2014)\citenamefont {Sun},
  \citenamefont {Zhang}, \citenamefont {Ren}, \citenamefont {Tu},\ and\
  \citenamefont {Liu}}]{Sun2014}%
  \BibitemOpen
  \bibfield  {author} {\bibinfo {author} {\bibfnamefont {Q.}~\bibnamefont
  {Sun}}, \bibinfo {author} {\bibfnamefont {X.}~\bibnamefont {Zhang}}, \bibinfo
  {author} {\bibfnamefont {Y.}~\bibnamefont {Ren}}, \bibinfo {author}
  {\bibfnamefont {J.}~\bibnamefont {Tu}},\ and\ \bibinfo {author}
  {\bibfnamefont {Q.}~\bibnamefont {Liu}},\ }\bibfield  {title} {\bibinfo
  {title} {Interfacial structure of \{10-12\} twin tip in deformed magnesium
  alloy},\ }\href {https://doi.org/10.1016/j.scriptamat.2014.07.012} {\bibfield
   {journal} {\bibinfo  {journal} {Scripta Materialia}\ }\textbf {\bibinfo
  {volume} {90-91}},\ \bibinfo {pages} {41} (\bibinfo {year}
  {2014})}\BibitemShut {NoStop}%
\bibitem [{\citenamefont {Howe}\ \emph {et~al.}(2009)\citenamefont {Howe},
  \citenamefont {Pond},\ and\ \citenamefont {Hirth}}]{Howe2009}%
  \BibitemOpen
  \bibfield  {author} {\bibinfo {author} {\bibfnamefont {J.~M.}\ \bibnamefont
  {Howe}}, \bibinfo {author} {\bibfnamefont {R.~C.}\ \bibnamefont {Pond}},\
  and\ \bibinfo {author} {\bibfnamefont {J.~P.}\ \bibnamefont {Hirth}},\ }\href
  {https://doi.org/10.1016/j.pmatsci.2009.04.001} {\bibinfo {title} {The role
  of disconnections in phase transformations}} (\bibinfo {year}
  {2009})\BibitemShut {NoStop}%
\bibitem [{\citenamefont {Xu}\ \emph {et~al.}(2013)\citenamefont {Xu},
  \citenamefont {Capolungo},\ and\ \citenamefont {Rodney}}]{Xu2013}%
  \BibitemOpen
  \bibfield  {author} {\bibinfo {author} {\bibfnamefont {B.}~\bibnamefont
  {Xu}}, \bibinfo {author} {\bibfnamefont {L.}~\bibnamefont {Capolungo}},\ and\
  \bibinfo {author} {\bibfnamefont {D.}~\bibnamefont {Rodney}},\ }\bibfield
  {title} {\bibinfo {title} {On the importance of prismatic/basal interfaces in
  the growth of twins in hexagonal close packed crystals},\ }\href
  {https://doi.org/10.1016/j.scriptamat.2013.02.023} {\bibfield  {journal}
  {\bibinfo  {journal} {Scripta Materialia}\ }\textbf {\bibinfo {volume}
  {68}},\ \bibinfo {pages} {901} (\bibinfo {year} {2013})}\BibitemShut
  {NoStop}%
\bibitem [{\citenamefont {Hu}\ \emph {et~al.}(2020{\natexlab{a}})\citenamefont
  {Hu}, \citenamefont {Turlo}, \citenamefont {Beyerlein}, \citenamefont
  {Mahajan}, \citenamefont {Lavernia}, \citenamefont {Schoenung},\ and\
  \citenamefont {Rupert}}]{hu2020disconnection}%
  \BibitemOpen
  \bibfield  {author} {\bibinfo {author} {\bibfnamefont {Y.}~\bibnamefont
  {Hu}}, \bibinfo {author} {\bibfnamefont {V.}~\bibnamefont {Turlo}}, \bibinfo
  {author} {\bibfnamefont {I.~J.}\ \bibnamefont {Beyerlein}}, \bibinfo {author}
  {\bibfnamefont {S.}~\bibnamefont {Mahajan}}, \bibinfo {author} {\bibfnamefont
  {E.~J.}\ \bibnamefont {Lavernia}}, \bibinfo {author} {\bibfnamefont {J.~M.}\
  \bibnamefont {Schoenung}},\ and\ \bibinfo {author} {\bibfnamefont {T.~J.}\
  \bibnamefont {Rupert}},\ }\bibfield  {title} {\bibinfo {title}
  {Disconnection-mediated twin embryo growth in mg},\ }\href@noop {} {\bibfield
   {journal} {\bibinfo  {journal} {Acta Materialia}\ }\textbf {\bibinfo
  {volume} {194}},\ \bibinfo {pages} {437} (\bibinfo {year}
  {2020}{\natexlab{a}})}\BibitemShut {NoStop}%
\bibitem [{\citenamefont {Huang}\ \emph {et~al.}(2021)\citenamefont {Huang},
  \citenamefont {Turlo}, \citenamefont {Wang}, \citenamefont {Chen},
  \citenamefont {Shen}, \citenamefont {Zhang}, \citenamefont {Beyerlein},\ and\
  \citenamefont {Rupert}}]{Huang2021}%
  \BibitemOpen
  \bibfield  {author} {\bibinfo {author} {\bibfnamefont {Z.}~\bibnamefont
  {Huang}}, \bibinfo {author} {\bibfnamefont {V.}~\bibnamefont {Turlo}},
  \bibinfo {author} {\bibfnamefont {X.}~\bibnamefont {Wang}}, \bibinfo {author}
  {\bibfnamefont {F.}~\bibnamefont {Chen}}, \bibinfo {author} {\bibfnamefont
  {Q.}~\bibnamefont {Shen}}, \bibinfo {author} {\bibfnamefont {L.}~\bibnamefont
  {Zhang}}, \bibinfo {author} {\bibfnamefont {I.~J.}\ \bibnamefont
  {Beyerlein}},\ and\ \bibinfo {author} {\bibfnamefont {T.~J.}\ \bibnamefont
  {Rupert}},\ }\bibfield  {title} {\bibinfo {title} {Dislocation-induced y
  segregation at basal-prismatic interfaces in mg},\ }\bibfield  {journal}
  {\bibinfo  {journal} {Computational Materials Science}\ }\textbf {\bibinfo
  {volume} {188}},\ \href {https://doi.org/10.1016/j.commatsci.2020.110241}
  {10.1016/j.commatsci.2020.110241} (\bibinfo {year} {2021})\BibitemShut
  {NoStop}%
\bibitem [{\citenamefont {Kumar}\ \emph {et~al.}(2015)\citenamefont {Kumar},
  \citenamefont {Wang},\ and\ \citenamefont {Tomé}}]{Kumar2015}%
  \BibitemOpen
  \bibfield  {author} {\bibinfo {author} {\bibfnamefont {A.}~\bibnamefont
  {Kumar}}, \bibinfo {author} {\bibfnamefont {J.}~\bibnamefont {Wang}},\ and\
  \bibinfo {author} {\bibfnamefont {C.~N.}\ \bibnamefont {Tomé}},\ }\bibfield
  {title} {\bibinfo {title} {First-principles study of energy and atomic
  solubility of twinning-associated boundaries in hexagonal metals},\
  }\bibfield  {journal} {\bibinfo  {journal} {Acta Materialia}\ }\textbf
  {\bibinfo {volume} {85}},\ \href
  {https://doi.org/10.1016/j.actamat.2014.11.015}
  {10.1016/j.actamat.2014.11.015} (\bibinfo {year} {2015})\BibitemShut
  {NoStop}%
\bibitem [{\citenamefont {Yu}\ \emph {et~al.}(2014{\natexlab{a}})\citenamefont
  {Yu}, \citenamefont {Wang}, \citenamefont {Jiang}, \citenamefont {McCabe},
  \citenamefont {Li},\ and\ \citenamefont {Tomé}}]{Yu2014}%
  \BibitemOpen
  \bibfield  {author} {\bibinfo {author} {\bibfnamefont {Q.}~\bibnamefont
  {Yu}}, \bibinfo {author} {\bibfnamefont {J.}~\bibnamefont {Wang}}, \bibinfo
  {author} {\bibfnamefont {Y.}~\bibnamefont {Jiang}}, \bibinfo {author}
  {\bibfnamefont {R.~J.}\ \bibnamefont {McCabe}}, \bibinfo {author}
  {\bibfnamefont {N.}~\bibnamefont {Li}},\ and\ \bibinfo {author}
  {\bibfnamefont {C.~N.}\ \bibnamefont {Tomé}},\ }\bibfield  {title} {\bibinfo
  {title} {Twin–twin interactions in magnesium},\ }\href
  {https://doi.org/10.1016/j.actamat.2014.05.030} {\bibfield  {journal}
  {\bibinfo  {journal} {Acta Materialia}\ }\textbf {\bibinfo {volume} {77}},\
  \bibinfo {pages} {28} (\bibinfo {year} {2014}{\natexlab{a}})}\BibitemShut
  {NoStop}%
\bibitem [{\citenamefont {Yu}\ \emph {et~al.}(2014{\natexlab{b}})\citenamefont
  {Yu}, \citenamefont {Wang}, \citenamefont {Jiang}, \citenamefont {McCabe},\
  and\ \citenamefont {Tomé}}]{Yu2014cozone}%
  \BibitemOpen
  \bibfield  {author} {\bibinfo {author} {\bibfnamefont {Q.}~\bibnamefont
  {Yu}}, \bibinfo {author} {\bibfnamefont {J.}~\bibnamefont {Wang}}, \bibinfo
  {author} {\bibfnamefont {Y.}~\bibnamefont {Jiang}}, \bibinfo {author}
  {\bibfnamefont {R.~J.}\ \bibnamefont {McCabe}},\ and\ \bibinfo {author}
  {\bibfnamefont {C.~N.}\ \bibnamefont {Tomé}},\ }\bibfield  {title} {\bibinfo
  {title} {Co-zone (1012) twin interaction in magnesium single crystal},\
  }\bibfield  {journal} {\bibinfo  {journal} {Materials Research Letters}\
  }\textbf {\bibinfo {volume} {2}},\ \href
  {https://doi.org/10.1080/21663831.2013.867291} {10.1080/21663831.2013.867291}
  (\bibinfo {year} {2014}{\natexlab{b}})\BibitemShut {NoStop}%
\bibitem [{\citenamefont {Kumar}\ \emph {et~al.}(2016)\citenamefont {Kumar},
  \citenamefont {Beyerlein}, \citenamefont {McCabe},\ and\ \citenamefont
  {Tomé}}]{ArulKumar2016}%
  \BibitemOpen
  \bibfield  {author} {\bibinfo {author} {\bibfnamefont {M.~A.}\ \bibnamefont
  {Kumar}}, \bibinfo {author} {\bibfnamefont {I.~J.}\ \bibnamefont
  {Beyerlein}}, \bibinfo {author} {\bibfnamefont {R.~J.}\ \bibnamefont
  {McCabe}},\ and\ \bibinfo {author} {\bibfnamefont {C.~N.}\ \bibnamefont
  {Tomé}},\ }\bibfield  {title} {\bibinfo {title} {Grain neighbour effects on
  twin transmission in hexagonal close-packed materials},\ }\href
  {https://doi.org/10.1038/ncomms13826} {\bibfield  {journal} {\bibinfo
  {journal} {Nature Communications}\ }\textbf {\bibinfo {volume} {7}},\
  \bibinfo {pages} {13826} (\bibinfo {year} {2016})}\BibitemShut {NoStop}%
\bibitem [{\citenamefont {Nie}\ \emph {et~al.}(2013)\citenamefont {Nie},
  \citenamefont {Zhu}, \citenamefont {Liu},\ and\ \citenamefont
  {Fang}}]{Nie2013}%
  \BibitemOpen
  \bibfield  {author} {\bibinfo {author} {\bibfnamefont {J.~F.}\ \bibnamefont
  {Nie}}, \bibinfo {author} {\bibfnamefont {Y.~M.}\ \bibnamefont {Zhu}},
  \bibinfo {author} {\bibfnamefont {J.~Z.}\ \bibnamefont {Liu}},\ and\ \bibinfo
  {author} {\bibfnamefont {X.~Y.}\ \bibnamefont {Fang}},\ }\bibfield  {title}
  {\bibinfo {title} {Periodic segregation of solute atoms in fully coherent
  twin boundaries},\ }\href {https://doi.org/10.1126/science.1229369}
  {\bibfield  {journal} {\bibinfo  {journal} {Science}\ }\textbf {\bibinfo
  {volume} {340}},\ \bibinfo {pages} {957} (\bibinfo {year}
  {2013})}\BibitemShut {NoStop}%
\bibitem [{\citenamefont {Ishii}\ \emph {et~al.}(2016)\citenamefont {Ishii},
  \citenamefont {Li},\ and\ \citenamefont {Ogata}}]{Ishii2016}%
  \BibitemOpen
  \bibfield  {author} {\bibinfo {author} {\bibfnamefont {A.}~\bibnamefont
  {Ishii}}, \bibinfo {author} {\bibfnamefont {J.}~\bibnamefont {Li}},\ and\
  \bibinfo {author} {\bibfnamefont {S.}~\bibnamefont {Ogata}},\ }\bibfield
  {title} {\bibinfo {title} {Shuffling-controlled versus strain-controlled
  deformation twinning: The case for hcp mg twin nucleation},\ }\bibfield
  {journal} {\bibinfo  {journal} {International Journal of Plasticity}\
  }\textbf {\bibinfo {volume} {82}},\ \href
  {https://doi.org/10.1016/j.ijplas.2016.01.019} {10.1016/j.ijplas.2016.01.019}
  (\bibinfo {year} {2016})\BibitemShut {NoStop}%
\bibitem [{\citenamefont {Hu}\ \emph {et~al.}(2020{\natexlab{b}})\citenamefont
  {Hu}, \citenamefont {Turlo}, \citenamefont {Beyerlein}, \citenamefont
  {Mahajan}, \citenamefont {Lavernia}, \citenamefont {Schoenung},\ and\
  \citenamefont {Rupert}}]{Hu2020}%
  \BibitemOpen
  \bibfield  {author} {\bibinfo {author} {\bibfnamefont {Y.}~\bibnamefont
  {Hu}}, \bibinfo {author} {\bibfnamefont {V.}~\bibnamefont {Turlo}}, \bibinfo
  {author} {\bibfnamefont {I.~J.}\ \bibnamefont {Beyerlein}}, \bibinfo {author}
  {\bibfnamefont {S.}~\bibnamefont {Mahajan}}, \bibinfo {author} {\bibfnamefont
  {E.~J.}\ \bibnamefont {Lavernia}}, \bibinfo {author} {\bibfnamefont {J.~M.}\
  \bibnamefont {Schoenung}},\ and\ \bibinfo {author} {\bibfnamefont {T.~J.}\
  \bibnamefont {Rupert}},\ }\bibfield  {title} {\bibinfo {title} {Embracing the
  chaos: Alloying adds stochasticity to twin embryo growth},\ }\href
  {https://doi.org/10.1103/PhysRevLett.125.205503} {\bibfield  {journal}
  {\bibinfo  {journal} {Physical Review Letters}\ }\textbf {\bibinfo {volume}
  {125}},\ \bibinfo {pages} {205503} (\bibinfo {year}
  {2020}{\natexlab{b}})}\BibitemShut {NoStop}%
\bibitem [{\citenamefont {Cheng}\ and\ \citenamefont
  {Ghosh}(2015)}]{Cheng2015}%
  \BibitemOpen
  \bibfield  {author} {\bibinfo {author} {\bibfnamefont {J.}~\bibnamefont
  {Cheng}}\ and\ \bibinfo {author} {\bibfnamefont {S.}~\bibnamefont {Ghosh}},\
  }\bibfield  {title} {\bibinfo {title} {A crystal plasticity fe model for
  deformation with twin nucleation in magnesium alloys},\ }\bibfield  {journal}
  {\bibinfo  {journal} {International Journal of Plasticity}\ }\textbf
  {\bibinfo {volume} {67}},\ \href
  {https://doi.org/10.1016/j.ijplas.2014.10.005} {10.1016/j.ijplas.2014.10.005}
  (\bibinfo {year} {2015})\BibitemShut {NoStop}%
\bibitem [{\citenamefont {Hollenweger}\ and\ \citenamefont
  {Kochmann}(2022)}]{Hollenweger2022}%
  \BibitemOpen
  \bibfield  {author} {\bibinfo {author} {\bibfnamefont {Y.}~\bibnamefont
  {Hollenweger}}\ and\ \bibinfo {author} {\bibfnamefont {D.~M.}\ \bibnamefont
  {Kochmann}},\ }\bibfield  {title} {\bibinfo {title} {An efficient
  temperature-dependent crystal plasticity framework for pure magnesium with
  emphasis on the competition between slip and twinning},\ }\bibfield
  {journal} {\bibinfo  {journal} {International Journal of Plasticity}\
  }\textbf {\bibinfo {volume} {159}},\ \href
  {https://doi.org/10.1016/j.ijplas.2022.103448} {10.1016/j.ijplas.2022.103448}
  (\bibinfo {year} {2022})\BibitemShut {NoStop}%
\bibitem [{\citenamefont {Chang}\ \emph {et~al.}(2017)\citenamefont {Chang},
  \citenamefont {Lloyd}, \citenamefont {Becker},\ and\ \citenamefont
  {Kochmann}}]{Chang2017}%
  \BibitemOpen
  \bibfield  {author} {\bibinfo {author} {\bibfnamefont {Y.}~\bibnamefont
  {Chang}}, \bibinfo {author} {\bibfnamefont {J.~T.}\ \bibnamefont {Lloyd}},
  \bibinfo {author} {\bibfnamefont {R.}~\bibnamefont {Becker}},\ and\ \bibinfo
  {author} {\bibfnamefont {D.~M.}\ \bibnamefont {Kochmann}},\ }\bibfield
  {title} {\bibinfo {title} {Modeling microstructure evolution in magnesium:
  Comparison of detailed and reduced-order kinematic models},\ }\bibfield
  {journal} {\bibinfo  {journal} {Mechanics of Materials}\ }\textbf {\bibinfo
  {volume} {108}},\ \href {https://doi.org/10.1016/j.mechmat.2017.02.007}
  {10.1016/j.mechmat.2017.02.007} (\bibinfo {year} {2017})\BibitemShut
  {NoStop}%
\bibitem [{\citenamefont {Liu}\ \emph {et~al.}(2018)\citenamefont {Liu},
  \citenamefont {Shanthraj}, \citenamefont {Diehl}, \citenamefont {Roters},
  \citenamefont {Dong}, \citenamefont {Dong}, \citenamefont {Ding},\ and\
  \citenamefont {Raabe}}]{Liu2018}%
  \BibitemOpen
  \bibfield  {author} {\bibinfo {author} {\bibfnamefont {C.}~\bibnamefont
  {Liu}}, \bibinfo {author} {\bibfnamefont {P.}~\bibnamefont {Shanthraj}},
  \bibinfo {author} {\bibfnamefont {M.}~\bibnamefont {Diehl}}, \bibinfo
  {author} {\bibfnamefont {F.}~\bibnamefont {Roters}}, \bibinfo {author}
  {\bibfnamefont {S.}~\bibnamefont {Dong}}, \bibinfo {author} {\bibfnamefont
  {J.}~\bibnamefont {Dong}}, \bibinfo {author} {\bibfnamefont {W.}~\bibnamefont
  {Ding}},\ and\ \bibinfo {author} {\bibfnamefont {D.}~\bibnamefont {Raabe}},\
  }\bibfield  {title} {\bibinfo {title} {An integrated crystal plasticity-phase
  field model for spatially resolved twin nucleation, propagation, and growth
  in hexagonal materials},\ }\bibfield  {journal} {\bibinfo  {journal}
  {International Journal of Plasticity}\ }\textbf {\bibinfo {volume} {106}},\
  \href {https://doi.org/10.1016/j.ijplas.2018.03.009}
  {10.1016/j.ijplas.2018.03.009} (\bibinfo {year} {2018})\BibitemShut {NoStop}%
\bibitem [{\citenamefont {Kondo}\ \emph {et~al.}(2014)\citenamefont {Kondo},
  \citenamefont {Tadano},\ and\ \citenamefont {Shizawa}}]{kondo2014phase}%
  \BibitemOpen
  \bibfield  {author} {\bibinfo {author} {\bibfnamefont {R.}~\bibnamefont
  {Kondo}}, \bibinfo {author} {\bibfnamefont {Y.}~\bibnamefont {Tadano}},\ and\
  \bibinfo {author} {\bibfnamefont {K.}~\bibnamefont {Shizawa}},\ }\bibfield
  {title} {\bibinfo {title} {A phase-field model of twinning and detwinning
  coupled with dislocation-based crystal plasticity for hcp metals},\
  }\href@noop {} {\bibfield  {journal} {\bibinfo  {journal} {Computational
  materials science}\ }\textbf {\bibinfo {volume} {95}},\ \bibinfo {pages}
  {672} (\bibinfo {year} {2014})}\BibitemShut {NoStop}%
\bibitem [{\citenamefont {Clayton}\ and\ \citenamefont
  {Knap}(2011)}]{Clayton2011}%
  \BibitemOpen
  \bibfield  {author} {\bibinfo {author} {\bibfnamefont {J.~D.}\ \bibnamefont
  {Clayton}}\ and\ \bibinfo {author} {\bibfnamefont {J.}~\bibnamefont {Knap}},\
  }\bibfield  {title} {\bibinfo {title} {A phase field model of deformation
  twinning: Nonlinear theory and numerical simulations},\ }\bibfield  {journal}
  {\bibinfo  {journal} {Physica D: Nonlinear Phenomena}\ }\textbf {\bibinfo
  {volume} {240}},\ \href {https://doi.org/10.1016/j.physd.2010.12.012}
  {10.1016/j.physd.2010.12.012} (\bibinfo {year} {2011})\BibitemShut {NoStop}%
\bibitem [{\citenamefont {Agrawal}\ and\ \citenamefont
  {Dayal}(2015)}]{Agrawal2015}%
  \BibitemOpen
  \bibfield  {author} {\bibinfo {author} {\bibfnamefont {V.}~\bibnamefont
  {Agrawal}}\ and\ \bibinfo {author} {\bibfnamefont {K.}~\bibnamefont
  {Dayal}},\ }\bibfield  {title} {\bibinfo {title} {A dynamic phase-field model
  for structural transformations and twinning: Regularized interfaces with
  transparent prescription of complex kinetics and nucleation. part i:
  Formulation and one-dimensional characterization},\ }\bibfield  {journal}
  {\bibinfo  {journal} {Journal of the Mechanics and Physics of Solids}\
  }\textbf {\bibinfo {volume} {85}},\ \href
  {https://doi.org/10.1016/j.jmps.2015.04.010} {10.1016/j.jmps.2015.04.010}
  (\bibinfo {year} {2015})\BibitemShut {NoStop}%
\bibitem [{\citenamefont {Amirian}\ \emph {et~al.}(2022)\citenamefont
  {Amirian}, \citenamefont {Jafarzadeh}, \citenamefont {Abali}, \citenamefont
  {Reali},\ and\ \citenamefont {Hogan}}]{Amirian2022}%
  \BibitemOpen
  \bibfield  {author} {\bibinfo {author} {\bibfnamefont {B.}~\bibnamefont
  {Amirian}}, \bibinfo {author} {\bibfnamefont {H.}~\bibnamefont {Jafarzadeh}},
  \bibinfo {author} {\bibfnamefont {B.~E.}\ \bibnamefont {Abali}}, \bibinfo
  {author} {\bibfnamefont {A.}~\bibnamefont {Reali}},\ and\ \bibinfo {author}
  {\bibfnamefont {J.~D.}\ \bibnamefont {Hogan}},\ }\bibfield  {title} {\bibinfo
  {title} {Phase-field approach to evolution and interaction of twins in single
  crystal magnesium},\ }\bibfield  {journal} {\bibinfo  {journal}
  {Computational Mechanics}\ }\textbf {\bibinfo {volume} {70}},\ \href
  {https://doi.org/10.1007/s00466-022-02209-3} {10.1007/s00466-022-02209-3}
  (\bibinfo {year} {2022})\BibitemShut {NoStop}%
\bibitem [{\citenamefont {Spearot}\ \emph {et~al.}(2020)\citenamefont
  {Spearot}, \citenamefont {Taupin}, \citenamefont {Dang},\ and\ \citenamefont
  {Capolungo}}]{spearot2020structure}%
  \BibitemOpen
  \bibfield  {author} {\bibinfo {author} {\bibfnamefont {D.~E.}\ \bibnamefont
  {Spearot}}, \bibinfo {author} {\bibfnamefont {V.}~\bibnamefont {Taupin}},
  \bibinfo {author} {\bibfnamefont {K.}~\bibnamefont {Dang}},\ and\ \bibinfo
  {author} {\bibfnamefont {L.}~\bibnamefont {Capolungo}},\ }\bibfield  {title}
  {\bibinfo {title} {Structure and kinetics of three-dimensional defects on the
  $\{10-12\}$ twin boundary in magnesium: Atomistic and phase-field
  simulations},\ }\href@noop {} {\bibfield  {journal} {\bibinfo  {journal}
  {Mechanics of Materials}\ }\textbf {\bibinfo {volume} {143}},\ \bibinfo
  {pages} {103314} (\bibinfo {year} {2020})}\BibitemShut {NoStop}%
\bibitem [{\citenamefont {Gong}\ \emph {et~al.}(2021)\citenamefont {Gong},
  \citenamefont {Graham}, \citenamefont {Taupin},\ and\ \citenamefont
  {Capolungo}}]{gong2021effects}%
  \BibitemOpen
  \bibfield  {author} {\bibinfo {author} {\bibfnamefont {M.}~\bibnamefont
  {Gong}}, \bibinfo {author} {\bibfnamefont {J.}~\bibnamefont {Graham}},
  \bibinfo {author} {\bibfnamefont {V.}~\bibnamefont {Taupin}},\ and\ \bibinfo
  {author} {\bibfnamefont {L.}~\bibnamefont {Capolungo}},\ }\bibfield  {title}
  {\bibinfo {title} {The effects of stress, temperature and facet structure on
  growth of $\{10-12\}$ twins in mg: A molecular dynamics and phase field
  study},\ }\href@noop {} {\bibfield  {journal} {\bibinfo  {journal} {Acta
  Materialia}\ }\textbf {\bibinfo {volume} {208}},\ \bibinfo {pages} {116603}
  (\bibinfo {year} {2021})}\BibitemShut {NoStop}%
\bibitem [{\citenamefont {Fang}\ \emph {et~al.}(2022)\citenamefont {Fang},
  \citenamefont {Xiao}, \citenamefont {Tan}, \citenamefont {Deng},
  \citenamefont {Wang},\ and\ \citenamefont {Mao}}]{Fang2022}%
  \BibitemOpen
  \bibfield  {author} {\bibinfo {author} {\bibfnamefont {Z.}~\bibnamefont
  {Fang}}, \bibinfo {author} {\bibfnamefont {J.}~\bibnamefont {Xiao}}, \bibinfo
  {author} {\bibfnamefont {S.}~\bibnamefont {Tan}}, \bibinfo {author}
  {\bibfnamefont {C.}~\bibnamefont {Deng}}, \bibinfo {author} {\bibfnamefont
  {G.}~\bibnamefont {Wang}},\ and\ \bibinfo {author} {\bibfnamefont {S.~X.}\
  \bibnamefont {Mao}},\ }\bibfield  {title} {\bibinfo {title} {Atomic-scale
  observation of dynamic grain boundary structural transformation during
  shear-mediated migration},\ }\bibfield  {journal} {\bibinfo  {journal}
  {Science Advances}\ }\textbf {\bibinfo {volume} {8}},\ \href
  {https://doi.org/10.1126/sciadv.abn3785} {10.1126/sciadv.abn3785} (\bibinfo
  {year} {2022})\BibitemShut {NoStop}%
\bibitem [{\citenamefont {Zhu}\ \emph {et~al.}(2019)\citenamefont {Zhu},
  \citenamefont {Cao}, \citenamefont {Wang}, \citenamefont {Deng},
  \citenamefont {Li}, \citenamefont {Zhang},\ and\ \citenamefont
  {Mao}}]{Zhu2019}%
  \BibitemOpen
  \bibfield  {author} {\bibinfo {author} {\bibfnamefont {Q.}~\bibnamefont
  {Zhu}}, \bibinfo {author} {\bibfnamefont {G.}~\bibnamefont {Cao}}, \bibinfo
  {author} {\bibfnamefont {J.}~\bibnamefont {Wang}}, \bibinfo {author}
  {\bibfnamefont {C.}~\bibnamefont {Deng}}, \bibinfo {author} {\bibfnamefont
  {J.}~\bibnamefont {Li}}, \bibinfo {author} {\bibfnamefont {Z.}~\bibnamefont
  {Zhang}},\ and\ \bibinfo {author} {\bibfnamefont {S.~X.}\ \bibnamefont
  {Mao}},\ }\bibfield  {title} {\bibinfo {title} {In situ atomistic observation
  of disconnection-mediated grain boundary migration},\ }\bibfield  {journal}
  {\bibinfo  {journal} {Nature Communications}\ }\textbf {\bibinfo {volume}
  {10}},\ \href {https://doi.org/10.1038/s41467-018-08031-x}
  {10.1038/s41467-018-08031-x} (\bibinfo {year} {2019})\BibitemShut {NoStop}%
\bibitem [{\citenamefont {Sternlicht}\ \emph {et~al.}(2019)\citenamefont
  {Sternlicht}, \citenamefont {Rheinheimer}, \citenamefont {Dunin-Borkowski},
  \citenamefont {Hoffmann},\ and\ \citenamefont {Kaplan}}]{Sternlicht2019}%
  \BibitemOpen
  \bibfield  {author} {\bibinfo {author} {\bibfnamefont {H.}~\bibnamefont
  {Sternlicht}}, \bibinfo {author} {\bibfnamefont {W.}~\bibnamefont
  {Rheinheimer}}, \bibinfo {author} {\bibfnamefont {R.~E.}\ \bibnamefont
  {Dunin-Borkowski}}, \bibinfo {author} {\bibfnamefont {M.~J.}\ \bibnamefont
  {Hoffmann}},\ and\ \bibinfo {author} {\bibfnamefont {W.~D.}\ \bibnamefont
  {Kaplan}},\ }\bibfield  {title} {\bibinfo {title} {Characterization of grain
  boundary disconnections in srtio3 part i: the dislocation component of grain
  boundary disconnections},\ }\bibfield  {journal} {\bibinfo  {journal}
  {Journal of Materials Science}\ }\textbf {\bibinfo {volume} {54}},\ \href
  {https://doi.org/10.1007/s10853-018-3096-4} {10.1007/s10853-018-3096-4}
  (\bibinfo {year} {2019})\BibitemShut {NoStop}%
\bibitem [{\citenamefont {Sternlicht}\ \emph {et~al.}(2022)\citenamefont
  {Sternlicht}, \citenamefont {McComb},\ and\ \citenamefont
  {Padture}}]{Sternlicht2022}%
  \BibitemOpen
  \bibfield  {author} {\bibinfo {author} {\bibfnamefont {H.}~\bibnamefont
  {Sternlicht}}, \bibinfo {author} {\bibfnamefont {D.~W.}\ \bibnamefont
  {McComb}},\ and\ \bibinfo {author} {\bibfnamefont {N.~P.}\ \bibnamefont
  {Padture}},\ }\bibfield  {title} {\bibinfo {title} {Interaction of ytterbium
  pyrosilicate environmental-barrier-coating ceramics with molten
  calcia-magnesia-aluminosilicate glass: Part ii, interfaces},\ }\bibfield
  {journal} {\bibinfo  {journal} {Acta Materialia}\ }\textbf {\bibinfo {volume}
  {241}},\ \href {https://doi.org/10.1016/j.actamat.2022.118359}
  {10.1016/j.actamat.2022.118359} (\bibinfo {year} {2022})\BibitemShut
  {NoStop}%
\bibitem [{\citenamefont {Chesser}\ \emph {et~al.}(2020)\citenamefont
  {Chesser}, \citenamefont {Yu}, \citenamefont {Deng}, \citenamefont {Holm},\
  and\ \citenamefont {Runnels}}]{Chesser2020}%
  \BibitemOpen
  \bibfield  {author} {\bibinfo {author} {\bibfnamefont {I.}~\bibnamefont
  {Chesser}}, \bibinfo {author} {\bibfnamefont {T.}~\bibnamefont {Yu}},
  \bibinfo {author} {\bibfnamefont {C.}~\bibnamefont {Deng}}, \bibinfo {author}
  {\bibfnamefont {E.}~\bibnamefont {Holm}},\ and\ \bibinfo {author}
  {\bibfnamefont {B.}~\bibnamefont {Runnels}},\ }\bibfield  {title} {\bibinfo
  {title} {A continuum thermodynamic framework for grain boundary motion},\
  }\bibfield  {journal} {\bibinfo  {journal} {Journal of the Mechanics and
  Physics of Solids}\ }\textbf {\bibinfo {volume} {137}},\ \href
  {https://doi.org/10.1016/j.jmps.2019.103827} {10.1016/j.jmps.2019.103827}
  (\bibinfo {year} {2020})\BibitemShut {NoStop}%
\bibitem [{\citenamefont {Guin}\ and\ \citenamefont
  {Kochmann}(2023)}]{Guin2023}%
  \BibitemOpen
  \bibfield  {author} {\bibinfo {author} {\bibfnamefont {L.}~\bibnamefont
  {Guin}}\ and\ \bibinfo {author} {\bibfnamefont {D.~M.}\ \bibnamefont
  {Kochmann}},\ }\bibfield  {title} {\bibinfo {title} {A phase-field model for
  ferroelectrics with general kinetics, part i: Model formulation},\ }\bibfield
   {journal} {\bibinfo  {journal} {Journal of the Mechanics and Physics of
  Solids}\ }\textbf {\bibinfo {volume} {176}},\ \href
  {https://doi.org/10.1016/j.jmps.2023.105301} {10.1016/j.jmps.2023.105301}
  (\bibinfo {year} {2023})\BibitemShut {NoStop}%
\bibitem [{\citenamefont {Gokuli}\ and\ \citenamefont
  {Runnels}(2021)}]{Gokuli2021}%
  \BibitemOpen
  \bibfield  {author} {\bibinfo {author} {\bibfnamefont {M.}~\bibnamefont
  {Gokuli}}\ and\ \bibinfo {author} {\bibfnamefont {B.}~\bibnamefont
  {Runnels}},\ }\bibfield  {title} {\bibinfo {title} {Multiphase field modeling
  of grain boundary migration mediated by emergent disconnections},\ }\bibfield
   {journal} {\bibinfo  {journal} {Acta Materialia}\ }\textbf {\bibinfo
  {volume} {217}},\ \href {https://doi.org/10.1016/j.actamat.2021.117149}
  {10.1016/j.actamat.2021.117149} (\bibinfo {year} {2021})\BibitemShut
  {NoStop}%
\bibitem [{\citenamefont {Plimpton}(1995)}]{Plimpton1995}%
  \BibitemOpen
  \bibfield  {author} {\bibinfo {author} {\bibfnamefont {S.}~\bibnamefont
  {Plimpton}},\ }\bibfield  {title} {\bibinfo {title} {Fast parallel algorithms
  for short-range molecular dynamics},\ }\href
  {https://doi.org/10.1006/jcph.1995.1039} {\bibfield  {journal} {\bibinfo
  {journal} {Journal of Computational Physics}\ }\textbf {\bibinfo {volume}
  {117}},\ \bibinfo {pages} {1} (\bibinfo {year} {1995})}\BibitemShut {NoStop}%
\bibitem [{\citenamefont {Kim}\ \emph {et~al.}(2009)\citenamefont {Kim},
  \citenamefont {Kim},\ and\ \citenamefont {Lee}}]{Kim2009}%
  \BibitemOpen
  \bibfield  {author} {\bibinfo {author} {\bibfnamefont {Y.-M.}\ \bibnamefont
  {Kim}}, \bibinfo {author} {\bibfnamefont {N.~J.}\ \bibnamefont {Kim}},\ and\
  \bibinfo {author} {\bibfnamefont {B.-J.}\ \bibnamefont {Lee}},\ }\bibfield
  {title} {\bibinfo {title} {Atomistic modeling of pure mg and mg–al
  systems},\ }\href {https://doi.org/10.1016/j.calphad.2009.07.004} {\bibfield
  {journal} {\bibinfo  {journal} {Calphad}\ }\textbf {\bibinfo {volume} {33}},\
  \bibinfo {pages} {650} (\bibinfo {year} {2009})}\BibitemShut {NoStop}%
\bibitem [{\citenamefont {Hu}\ and\ \citenamefont
  {Kochmann}(2023)}]{hu2023atomistic}%
  \BibitemOpen
  \bibfield  {author} {\bibinfo {author} {\bibfnamefont {Y.}~\bibnamefont
  {Hu}}\ and\ \bibinfo {author} {\bibfnamefont {D.~M.}\ \bibnamefont
  {Kochmann}},\ }\bibfield  {title} {\bibinfo {title} {Atomistic insight into
  three-dimensional twin embryo growth in mg alloys},\ }\href@noop {}
  {\bibfield  {journal} {\bibinfo  {journal} {Journal of Materials Science}\
  }\textbf {\bibinfo {volume} {58}},\ \bibinfo {pages} {3972} (\bibinfo {year}
  {2023})}\BibitemShut {NoStop}%
\bibitem [{\citenamefont {Huang}\ and\ \citenamefont
  {Nie}(2021)}]{Huang2021Hydrogen}%
  \BibitemOpen
  \bibfield  {author} {\bibinfo {author} {\bibfnamefont {Z.}~\bibnamefont
  {Huang}}\ and\ \bibinfo {author} {\bibfnamefont {J.~F.}\ \bibnamefont
  {Nie}},\ }\bibfield  {title} {\bibinfo {title} {Interaction between hydrogen
  and solute atoms in \{10-12\} twin boundary and its impact on boundary
  cohesion in magnesium},\ }\bibfield  {journal} {\bibinfo  {journal} {Acta
  Materialia}\ }\textbf {\bibinfo {volume} {214}},\ \href
  {https://doi.org/10.1016/j.actamat.2021.117009}
  {10.1016/j.actamat.2021.117009} (\bibinfo {year} {2021})\BibitemShut
  {NoStop}%
\bibitem [{\citenamefont {Wang}\ and\ \citenamefont
  {Beyerlein}(2012)}]{Wang2012}%
  \BibitemOpen
  \bibfield  {author} {\bibinfo {author} {\bibfnamefont {J.}~\bibnamefont
  {Wang}}\ and\ \bibinfo {author} {\bibfnamefont {I.~J.}\ \bibnamefont
  {Beyerlein}},\ }\bibfield  {title} {\bibinfo {title} {Atomic structures of
  symmetric tilt grain boundaries in hexagonal close packed (hcp) crystals},\
  }\bibfield  {journal} {\bibinfo  {journal} {Modelling and Simulation in
  Materials Science and Engineering}\ }\textbf {\bibinfo {volume} {20}},\ \href
  {https://doi.org/10.1088/0965-0393/20/2/024002}
  {10.1088/0965-0393/20/2/024002} (\bibinfo {year} {2012})\BibitemShut
  {NoStop}%
\bibitem [{\citenamefont {Jain}\ \emph {et~al.}(2013)\citenamefont {Jain},
  \citenamefont {Ong}, \citenamefont {Hautier}, \citenamefont {Chen},
  \citenamefont {Richards}, \citenamefont {Dacek}, \citenamefont {Cholia},
  \citenamefont {Gunter}, \citenamefont {Skinner}, \citenamefont {Ceder},\ and\
  \citenamefont {Persson}}]{Jain2013}%
  \BibitemOpen
  \bibfield  {author} {\bibinfo {author} {\bibfnamefont {A.}~\bibnamefont
  {Jain}}, \bibinfo {author} {\bibfnamefont {S.~P.}\ \bibnamefont {Ong}},
  \bibinfo {author} {\bibfnamefont {G.}~\bibnamefont {Hautier}}, \bibinfo
  {author} {\bibfnamefont {W.}~\bibnamefont {Chen}}, \bibinfo {author}
  {\bibfnamefont {W.~D.}\ \bibnamefont {Richards}}, \bibinfo {author}
  {\bibfnamefont {S.}~\bibnamefont {Dacek}}, \bibinfo {author} {\bibfnamefont
  {S.}~\bibnamefont {Cholia}}, \bibinfo {author} {\bibfnamefont
  {D.}~\bibnamefont {Gunter}}, \bibinfo {author} {\bibfnamefont
  {D.}~\bibnamefont {Skinner}}, \bibinfo {author} {\bibfnamefont
  {G.}~\bibnamefont {Ceder}},\ and\ \bibinfo {author} {\bibfnamefont {K.~A.}\
  \bibnamefont {Persson}},\ }\href {https://doi.org/10.1063/1.4812323}
  {\bibinfo {title} {Commentary: The materials project: A materials genome
  approach to accelerating materials innovation}} (\bibinfo {year}
  {2013})\BibitemShut {NoStop}%
\bibitem [{\citenamefont {Yang}\ \emph {et~al.}(2021)\citenamefont {Yang},
  \citenamefont {Meng},\ and\ \citenamefont {Cao}}]{Yang2021}%
  \BibitemOpen
  \bibfield  {author} {\bibinfo {author} {\bibfnamefont {X.}~\bibnamefont
  {Yang}}, \bibinfo {author} {\bibfnamefont {Z.}~\bibnamefont {Meng}},\ and\
  \bibinfo {author} {\bibfnamefont {H.}~\bibnamefont {Cao}},\ }\bibfield
  {title} {\bibinfo {title} {First-principles calculations to investigate the
  third-order elastic constants and mechanical properties of mg, be, ti, zn,
  zr, and cd},\ }\bibfield  {journal} {\bibinfo  {journal} {Advances in
  Materials Science and Engineering}\ }\textbf {\bibinfo {volume} {2021}},\
  \href {https://doi.org/10.1155/2021/8726250} {10.1155/2021/8726250} (\bibinfo
  {year} {2021})\BibitemShut {NoStop}%
\bibitem [{\citenamefont {Slutsky}\ and\ \citenamefont
  {Garland}(1957)}]{Slutsky1957}%
  \BibitemOpen
  \bibfield  {author} {\bibinfo {author} {\bibfnamefont {L.~J.}\ \bibnamefont
  {Slutsky}}\ and\ \bibinfo {author} {\bibfnamefont {C.~W.}\ \bibnamefont
  {Garland}},\ }\bibfield  {title} {\bibinfo {title} {Elastic constants of
  magnesium from 4.2°k to 300°k},\ }\bibfield  {journal} {\bibinfo  {journal}
  {Physical Review}\ }\textbf {\bibinfo {volume} {107}},\ \href
  {https://doi.org/10.1103/PhysRev.107.972} {10.1103/PhysRev.107.972} (\bibinfo
  {year} {1957})\BibitemShut {NoStop}%
\bibitem [{\citenamefont {Liu}\ \emph {et~al.}(1996)\citenamefont {Liu},
  \citenamefont {Adams}, \citenamefont {Ercolessi},\ and\ \citenamefont
  {Moriarty}}]{Liu1996}%
  \BibitemOpen
  \bibfield  {author} {\bibinfo {author} {\bibfnamefont {X.~Y.}\ \bibnamefont
  {Liu}}, \bibinfo {author} {\bibfnamefont {J.~B.}\ \bibnamefont {Adams}},
  \bibinfo {author} {\bibfnamefont {F.}~\bibnamefont {Ercolessi}},\ and\
  \bibinfo {author} {\bibfnamefont {J.~A.}\ \bibnamefont {Moriarty}},\
  }\bibfield  {title} {\bibinfo {title} {Eam potential for magnesium from
  quantum mechanical forces},\ }\bibfield  {journal} {\bibinfo  {journal}
  {Modelling and Simulation in Materials Science and Engineering}\ }\textbf
  {\bibinfo {volume} {4}},\ \href {https://doi.org/10.1088/0965-0393/4/3/004}
  {10.1088/0965-0393/4/3/004} (\bibinfo {year} {1996})\BibitemShut {NoStop}%
\bibitem [{\citenamefont {Stukowski}(2010)}]{Stukowski2010}%
  \BibitemOpen
  \bibfield  {author} {\bibinfo {author} {\bibfnamefont {A.}~\bibnamefont
  {Stukowski}},\ }\bibfield  {title} {\bibinfo {title} {Visualization and
  analysis of atomistic simulation data with ovito–the open visualization
  tool},\ }\href {https://doi.org/10.1088/0965-0393/18/1/015012} {\bibfield
  {journal} {\bibinfo  {journal} {Modelling and Simulation in Materials Science
  and Engineering}\ }\textbf {\bibinfo {volume} {18}},\ \bibinfo {pages}
  {015012} (\bibinfo {year} {2010})}\BibitemShut {NoStop}%
\bibitem [{\citenamefont {Larsen}\ \emph {et~al.}(2016)\citenamefont {Larsen},
  \citenamefont {Schmidt},\ and\ \citenamefont {Schiøtz}}]{Larsen2016}%
  \BibitemOpen
  \bibfield  {author} {\bibinfo {author} {\bibfnamefont {P.~M.}\ \bibnamefont
  {Larsen}}, \bibinfo {author} {\bibfnamefont {S.}~\bibnamefont {Schmidt}},\
  and\ \bibinfo {author} {\bibfnamefont {J.}~\bibnamefont {Schiøtz}},\
  }\bibfield  {title} {\bibinfo {title} {Robust structural identification via
  polyhedral template matching},\ }\href
  {https://doi.org/10.1088/0965-0393/24/5/055007} {\bibfield  {journal}
  {\bibinfo  {journal} {Modelling and Simulation in Materials Science and
  Engineering}\ }\textbf {\bibinfo {volume} {24}},\ \bibinfo {pages} {055007}
  (\bibinfo {year} {2016})}\BibitemShut {NoStop}%
\bibitem [{\citenamefont {Abeyaratne}\ and\ \citenamefont
  {Knowles}(1990)}]{Abeyaratne1990}%
  \BibitemOpen
  \bibfield  {author} {\bibinfo {author} {\bibfnamefont {R.}~\bibnamefont
  {Abeyaratne}}\ and\ \bibinfo {author} {\bibfnamefont {J.~K.}\ \bibnamefont
  {Knowles}},\ }\bibfield  {title} {\bibinfo {title} {On the driving traction
  acting on a surface of strain discontinuity in a continuum},\ }\href
  {https://doi.org/10.1016/0022-5096(90)90003-m} {\bibfield  {journal}
  {\bibinfo  {journal} {Journal of the Mechanics and Physics of Solids}\
  }\textbf {\bibinfo {volume} {38}},\ \bibinfo {pages} {345} (\bibinfo {year}
  {1990})}\BibitemShut {NoStop}%
\bibitem [{\citenamefont {Shimizu}\ \emph {et~al.}(2007)\citenamefont
  {Shimizu}, \citenamefont {Ogata},\ and\ \citenamefont {Li}}]{Shimizu2007}%
  \BibitemOpen
  \bibfield  {author} {\bibinfo {author} {\bibfnamefont {F.}~\bibnamefont
  {Shimizu}}, \bibinfo {author} {\bibfnamefont {S.}~\bibnamefont {Ogata}},\
  and\ \bibinfo {author} {\bibfnamefont {J.}~\bibnamefont {Li}},\ }\bibfield
  {title} {\bibinfo {title} {Theory of shear banding in metallic glasses and
  molecular dynamics calculations},\ }\bibfield  {journal} {\bibinfo  {journal}
  {Materials Transactions}\ }\textbf {\bibinfo {volume} {48}},\ \href
  {https://doi.org/10.2320/matertrans.MJ200769} {10.2320/matertrans.MJ200769}
  (\bibinfo {year} {2007})\BibitemShut {NoStop}%
\bibitem [{\citenamefont {Falk}\ and\ \citenamefont {Langer}(1998)}]{Falk1998}%
  \BibitemOpen
  \bibfield  {author} {\bibinfo {author} {\bibfnamefont {M.~L.}\ \bibnamefont
  {Falk}}\ and\ \bibinfo {author} {\bibfnamefont {J.~S.}\ \bibnamefont
  {Langer}},\ }\bibfield  {title} {\bibinfo {title} {Dynamics of viscoplastic
  deformation in amorphous solids},\ }\bibfield  {journal} {\bibinfo  {journal}
  {Physical Review E - Statistical Physics, Plasmas, Fluids, and Related
  Interdisciplinary Topics}\ }\textbf {\bibinfo {volume} {57}},\ \href
  {https://doi.org/10.1103/PhysRevE.57.7192} {10.1103/PhysRevE.57.7192}
  (\bibinfo {year} {1998})\BibitemShut {NoStop}%
\bibitem [{\citenamefont {Barrett}\ and\ \citenamefont
  {Kadiri}(2014)}]{Barrett2014GB}%
  \BibitemOpen
  \bibfield  {author} {\bibinfo {author} {\bibfnamefont {C.~D.}\ \bibnamefont
  {Barrett}}\ and\ \bibinfo {author} {\bibfnamefont {H.~E.}\ \bibnamefont
  {Kadiri}},\ }\bibfield  {title} {\bibinfo {title} {The roles of grain
  boundary dislocations and disclinations in the nucleation of \{10-12\}
  twinning},\ }\bibfield  {journal} {\bibinfo  {journal} {Acta Materialia}\
  }\textbf {\bibinfo {volume} {63}},\ \href
  {https://doi.org/10.1016/j.actamat.2013.09.012}
  {10.1016/j.actamat.2013.09.012} (\bibinfo {year} {2014})\BibitemShut
  {NoStop}%
\bibitem [{\citenamefont {Ortiz}\ and\ \citenamefont
  {Repetto}(1999)}]{Ortiz1999}%
  \BibitemOpen
  \bibfield  {author} {\bibinfo {author} {\bibfnamefont {M.}~\bibnamefont
  {Ortiz}}\ and\ \bibinfo {author} {\bibfnamefont {E.~A.}\ \bibnamefont
  {Repetto}},\ }\bibfield  {title} {\bibinfo {title} {Nonconvex energy
  minimization and dislocation structures in ductile single crystals},\
  }\bibfield  {journal} {\bibinfo  {journal} {Journal of the Mechanics and
  Physics of Solids}\ }\textbf {\bibinfo {volume} {47}},\ \href
  {https://doi.org/10.1016/S0022-5096(97)00096-3}
  {10.1016/S0022-5096(97)00096-3} (\bibinfo {year} {1999})\BibitemShut
  {NoStop}%
\bibitem [{\citenamefont {Hackl}\ and\ \citenamefont
  {Kochmann}(2008)}]{Hackl2008}%
  \BibitemOpen
  \bibfield  {author} {\bibinfo {author} {\bibfnamefont {K.}~\bibnamefont
  {Hackl}}\ and\ \bibinfo {author} {\bibfnamefont {D.~M.}\ \bibnamefont
  {Kochmann}},\ }\bibfield  {title} {\bibinfo {title} {Relaxed potentials and
  evolution equations for inelastic microstructures}\ }(\bibinfo {year}
  {2008})\BibitemShut {NoStop}%
\bibitem [{\citenamefont {Carstensen}\ \emph {et~al.}(2002)\citenamefont
  {Carstensen}, \citenamefont {Hackl},\ and\ \citenamefont
  {Mielke}}]{Carstensen2002}%
  \BibitemOpen
  \bibfield  {author} {\bibinfo {author} {\bibfnamefont {C.}~\bibnamefont
  {Carstensen}}, \bibinfo {author} {\bibfnamefont {K.}~\bibnamefont {Hackl}},\
  and\ \bibinfo {author} {\bibfnamefont {A.}~\bibnamefont {Mielke}},\
  }\bibfield  {title} {\bibinfo {title} {Non-convex potentials and
  microstructures in finite-strain plasticity},\ }\bibfield  {journal}
  {\bibinfo  {journal} {Proceedings of the Royal Society A: Mathematical,
  Physical and Engineering Sciences}\ }\textbf {\bibinfo {volume} {458}},\
  \href {https://doi.org/10.1098/rspa.2001.0864} {10.1098/rspa.2001.0864}
  (\bibinfo {year} {2002})\BibitemShut {NoStop}%
\bibitem [{\citenamefont {Roubicek}(2010)}]{Tomas2010}%
  \BibitemOpen
  \bibfield  {author} {\bibinfo {author} {\bibfnamefont {T.}~\bibnamefont
  {Roubicek}},\ }\bibfield  {title} {\bibinfo {title} {Thermodynamics of
  rate-independent processes in viscous solids at small strains},\ }\bibfield
  {journal} {\bibinfo  {journal} {SIAM Journal on Mathematical Analysis}\
  }\textbf {\bibinfo {volume} {42}},\ \href {https://doi.org/10.1137/080729992}
  {10.1137/080729992} (\bibinfo {year} {2010})\BibitemShut {NoStop}%
\bibitem [{\citenamefont {Roubicek}(2009)}]{Roubicek2009}%
  \BibitemOpen
  \bibfield  {author} {\bibinfo {author} {\bibfnamefont {T.}~\bibnamefont
  {Roubicek}},\ }\bibfield  {title} {\bibinfo {title} {Rate-independent
  processes in viscous solids at small strains},\ }\bibfield  {journal}
  {\bibinfo  {journal} {Mathematical Methods in the Applied Sciences}\ }\textbf
  {\bibinfo {volume} {32}},\ \href {https://doi.org/10.1002/mma.1069}
  {10.1002/mma.1069} (\bibinfo {year} {2009})\BibitemShut {NoStop}%
\bibitem [{\citenamefont {Bugas}\ and\ \citenamefont
  {Runnels}(2024)}]{bugas2024grain}%
  \BibitemOpen
  \bibfield  {author} {\bibinfo {author} {\bibfnamefont {D.}~\bibnamefont
  {Bugas}}\ and\ \bibinfo {author} {\bibfnamefont {B.}~\bibnamefont
  {Runnels}},\ }\bibfield  {title} {\bibinfo {title} {Grain boundary network
  plasticity: Reduced-order modeling of deformation-driven shear-coupled
  microstructure evolution},\ }\href@noop {} {\bibfield  {journal} {\bibinfo
  {journal} {Journal of the Mechanics and Physics of Solids}\ }\textbf
  {\bibinfo {volume} {184}},\ \bibinfo {pages} {105541} (\bibinfo {year}
  {2024})}\BibitemShut {NoStop}%
\bibitem [{\citenamefont {Moelans}\ \emph
  {et~al.}(2008{\natexlab{a}})\citenamefont {Moelans}, \citenamefont
  {Blanpain},\ and\ \citenamefont {Wollants}}]{Moelans2008GB}%
  \BibitemOpen
  \bibfield  {author} {\bibinfo {author} {\bibfnamefont {N.}~\bibnamefont
  {Moelans}}, \bibinfo {author} {\bibfnamefont {B.}~\bibnamefont {Blanpain}},\
  and\ \bibinfo {author} {\bibfnamefont {P.}~\bibnamefont {Wollants}},\
  }\bibfield  {title} {\bibinfo {title} {Quantitative analysis of grain
  boundary properties in a generalized phase field model for grain growth in
  anisotropic systems},\ }\bibfield  {journal} {\bibinfo  {journal} {Physical
  Review B - Condensed Matter and Materials Physics}\ }\textbf {\bibinfo
  {volume} {78}},\ \href {https://doi.org/10.1103/PhysRevB.78.024113}
  {10.1103/PhysRevB.78.024113} (\bibinfo {year}
  {2008}{\natexlab{a}})\BibitemShut {NoStop}%
\bibitem [{\citenamefont {Moelans}\ \emph
  {et~al.}(2008{\natexlab{b}})\citenamefont {Moelans}, \citenamefont
  {Blanpain},\ and\ \citenamefont {Wollants}}]{Moelans2008}%
  \BibitemOpen
  \bibfield  {author} {\bibinfo {author} {\bibfnamefont {N.}~\bibnamefont
  {Moelans}}, \bibinfo {author} {\bibfnamefont {B.}~\bibnamefont {Blanpain}},\
  and\ \bibinfo {author} {\bibfnamefont {P.}~\bibnamefont {Wollants}},\
  }\bibfield  {title} {\bibinfo {title} {Quantitative phase-field approach for
  simulating grain growth in anisotropic systems with arbitrary inclination and
  misorientation dependence},\ }\bibfield  {journal} {\bibinfo  {journal}
  {Physical Review Letters}\ }\textbf {\bibinfo {volume} {101}},\ \href
  {https://doi.org/10.1103/PhysRevLett.101.025502}
  {10.1103/PhysRevLett.101.025502} (\bibinfo {year}
  {2008}{\natexlab{b}})\BibitemShut {NoStop}%
\bibitem [{\citenamefont {Sutton}\ and\ \citenamefont
  {Balluffi}(1995)}]{Sutton1995}%
  \BibitemOpen
  \bibfield  {author} {\bibinfo {author} {\bibfnamefont {A.}~\bibnamefont
  {Sutton}}\ and\ \bibinfo {author} {\bibfnamefont {R.}~\bibnamefont
  {Balluffi}},\ }\bibfield  {title} {\bibinfo {title} {Interfaces in
  crystalline materials clarendon},\ }\href@noop {} {\bibfield  {journal}
  {\bibinfo  {journal} {Monographs on the physics and chemistry of materials}\
  } (\bibinfo {year} {1995})}\BibitemShut {NoStop}%
\bibitem [{\citenamefont {Ribot}\ \emph {et~al.}(2019)\citenamefont {Ribot},
  \citenamefont {Agrawal},\ and\ \citenamefont {Runnels}}]{Ribot2019}%
  \BibitemOpen
  \bibfield  {author} {\bibinfo {author} {\bibfnamefont {J.~G.}\ \bibnamefont
  {Ribot}}, \bibinfo {author} {\bibfnamefont {V.}~\bibnamefont {Agrawal}},\
  and\ \bibinfo {author} {\bibfnamefont {B.}~\bibnamefont {Runnels}},\
  }\bibfield  {title} {\bibinfo {title} {A new approach for phase field
  modeling of grain boundaries with strongly nonconvex energy},\ }\bibfield
  {journal} {\bibinfo  {journal} {Modelling and Simulation in Materials Science
  and Engineering}\ }\textbf {\bibinfo {volume} {27}},\ \href
  {https://doi.org/10.1088/1361-651X/ab47a0} {10.1088/1361-651X/ab47a0}
  (\bibinfo {year} {2019})\BibitemShut {NoStop}%
\bibitem [{\citenamefont {Rockafellar}(1997)}]{Rockafellar1997}%
  \BibitemOpen
  \bibfield  {author} {\bibinfo {author} {\bibfnamefont {R.}~\bibnamefont
  {Rockafellar}},\ }\bibfield  {title} {\bibinfo {title} {Book - convex
  analysis},\ }\href@noop {} {\bibfield  {journal} {\bibinfo  {journal}
  {Princeton University Press}\ } (\bibinfo {year} {1997})}\BibitemShut
  {NoStop}%
\bibitem [{\citenamefont {Upmanyu}\ \emph {et~al.}(1999)\citenamefont
  {Upmanyu}, \citenamefont {Srolovitz}, \citenamefont {Shvindlerman},\ and\
  \citenamefont {Gottstein}}]{Upmanyu1999}%
  \BibitemOpen
  \bibfield  {author} {\bibinfo {author} {\bibfnamefont {M.}~\bibnamefont
  {Upmanyu}}, \bibinfo {author} {\bibfnamefont {D.~J.}\ \bibnamefont
  {Srolovitz}}, \bibinfo {author} {\bibfnamefont {L.~S.}\ \bibnamefont
  {Shvindlerman}},\ and\ \bibinfo {author} {\bibfnamefont {G.}~\bibnamefont
  {Gottstein}},\ }\bibfield  {title} {\bibinfo {title} {Misorientation
  dependence of intrinsic grain boundary mobility: simulation and experiment},\
  }\bibfield  {journal} {\bibinfo  {journal} {Acta Materialia}\ }\textbf
  {\bibinfo {volume} {47}},\ \href
  {https://doi.org/10.1016/S1359-6454(99)00240-2}
  {10.1016/S1359-6454(99)00240-2} (\bibinfo {year} {1999})\BibitemShut
  {NoStop}%
\bibitem [{\citenamefont {Gottstein}\ and\ \citenamefont
  {Schwarzer}(1992)}]{Gottstein1992}%
  \BibitemOpen
  \bibfield  {author} {\bibinfo {author} {\bibfnamefont {G.}~\bibnamefont
  {Gottstein}}\ and\ \bibinfo {author} {\bibfnamefont {F.}~\bibnamefont
  {Schwarzer}},\ }\bibfield  {title} {\bibinfo {title} {On the orientation
  dependence of grain boundary energy and mobility},\ }\bibfield  {journal}
  {\bibinfo  {journal} {Materials Science Forum}\ }\textbf {\bibinfo {volume}
  {94-96}},\ \href {https://doi.org/10.4028/www.scientific.net/msf.94-96.187}
  {10.4028/www.scientific.net/msf.94-96.187} (\bibinfo {year}
  {1992})\BibitemShut {NoStop}%
\bibitem [{\citenamefont {Molodov}\ \emph {et~al.}(2022)\citenamefont
  {Molodov}, \citenamefont {Shvindlerman},\ and\ \citenamefont
  {Gottstein}}]{Molodov2022}%
  \BibitemOpen
  \bibfield  {author} {\bibinfo {author} {\bibfnamefont {D.~A.}\ \bibnamefont
  {Molodov}}, \bibinfo {author} {\bibfnamefont {L.~S.}\ \bibnamefont
  {Shvindlerman}},\ and\ \bibinfo {author} {\bibfnamefont {G.}~\bibnamefont
  {Gottstein}},\ }\bibfield  {title} {\bibinfo {title} {Impact of grain
  boundary character on grain boundary kinetics},\ }\bibfield  {journal}
  {\bibinfo  {journal} {International Journal of Materials Research}\ }\textbf
  {\bibinfo {volume} {94}},\ \href {https://doi.org/10.1515/ijmr-2003-0203}
  {10.1515/ijmr-2003-0203} (\bibinfo {year} {2022})\BibitemShut {NoStop}%
\bibitem [{\citenamefont {Sun}\ and\ \citenamefont
  {Beckermann}(2007)}]{Sun2007}%
  \BibitemOpen
  \bibfield  {author} {\bibinfo {author} {\bibfnamefont {Y.}~\bibnamefont
  {Sun}}\ and\ \bibinfo {author} {\bibfnamefont {C.}~\bibnamefont
  {Beckermann}},\ }\bibfield  {title} {\bibinfo {title} {Sharp interface
  tracking using the phase-field equation},\ }\bibfield  {journal} {\bibinfo
  {journal} {Journal of Computational Physics}\ }\textbf {\bibinfo {volume}
  {220}},\ \href {https://doi.org/10.1016/j.jcp.2006.05.025}
  {10.1016/j.jcp.2006.05.025} (\bibinfo {year} {2007})\BibitemShut {NoStop}%
\bibitem [{\citenamefont {Deiterding}(2011)}]{Deiterding2011}%
  \BibitemOpen
  \bibfield  {author} {\bibinfo {author} {\bibfnamefont {R.}~\bibnamefont
  {Deiterding}},\ }\bibfield  {title} {\bibinfo {title} {Block-structured
  adaptive mesh refinement - theory, implementation and application},\ }\href
  {https://doi.org/10.1051/proc/201134002} {\bibfield  {journal} {\bibinfo
  {journal} {ESAIM: Proceedings}\ }\textbf {\bibinfo {volume} {34}},\ \bibinfo
  {pages} {97} (\bibinfo {year} {2011})}\BibitemShut {NoStop}%
\bibitem [{\citenamefont {Zhang}\ \emph {et~al.}(2019)\citenamefont {Zhang},
  \citenamefont {Almgren}, \citenamefont {Beckner}, \citenamefont {Bell},
  \citenamefont {Blaschke}, \citenamefont {Chan}, \citenamefont {Day},
  \citenamefont {Friesen}, \citenamefont {Gott}, \citenamefont {Graves},
  \citenamefont {Katz}, \citenamefont {Myers}, \citenamefont {Nguyen},
  \citenamefont {Nonaka}, \citenamefont {Rosso}, \citenamefont {Williams},\
  and\ \citenamefont {Zingale}}]{Zhang2019}%
  \BibitemOpen
  \bibfield  {author} {\bibinfo {author} {\bibfnamefont {W.}~\bibnamefont
  {Zhang}}, \bibinfo {author} {\bibfnamefont {A.}~\bibnamefont {Almgren}},
  \bibinfo {author} {\bibfnamefont {V.}~\bibnamefont {Beckner}}, \bibinfo
  {author} {\bibfnamefont {J.}~\bibnamefont {Bell}}, \bibinfo {author}
  {\bibfnamefont {J.}~\bibnamefont {Blaschke}}, \bibinfo {author}
  {\bibfnamefont {C.}~\bibnamefont {Chan}}, \bibinfo {author} {\bibfnamefont
  {M.}~\bibnamefont {Day}}, \bibinfo {author} {\bibfnamefont {B.}~\bibnamefont
  {Friesen}}, \bibinfo {author} {\bibfnamefont {K.}~\bibnamefont {Gott}},
  \bibinfo {author} {\bibfnamefont {D.}~\bibnamefont {Graves}}, \bibinfo
  {author} {\bibfnamefont {M.}~\bibnamefont {Katz}}, \bibinfo {author}
  {\bibfnamefont {A.}~\bibnamefont {Myers}}, \bibinfo {author} {\bibfnamefont
  {T.}~\bibnamefont {Nguyen}}, \bibinfo {author} {\bibfnamefont
  {A.}~\bibnamefont {Nonaka}}, \bibinfo {author} {\bibfnamefont
  {M.}~\bibnamefont {Rosso}}, \bibinfo {author} {\bibfnamefont
  {S.}~\bibnamefont {Williams}},\ and\ \bibinfo {author} {\bibfnamefont
  {M.}~\bibnamefont {Zingale}},\ }\bibfield  {title} {\bibinfo {title} {Amrex:
  a framework for block-structured adaptive mesh refinement},\ }\bibfield
  {journal} {\bibinfo  {journal} {Journal of Open Source Software}\ }\textbf
  {\bibinfo {volume} {4}},\ \href {https://doi.org/10.21105/joss.01370}
  {10.21105/joss.01370} (\bibinfo {year} {2019})\BibitemShut {NoStop}%
\bibitem [{\citenamefont {Runnels}\ and\ \citenamefont
  {Agrawal}(2020)}]{Runnels2020}%
  \BibitemOpen
  \bibfield  {author} {\bibinfo {author} {\bibfnamefont {B.}~\bibnamefont
  {Runnels}}\ and\ \bibinfo {author} {\bibfnamefont {V.}~\bibnamefont
  {Agrawal}},\ }\bibfield  {title} {\bibinfo {title} {Phase field
  disconnections: A continuum method for disconnection-mediated grain boundary
  motion},\ }\bibfield  {journal} {\bibinfo  {journal} {Scripta Materialia}\
  }\textbf {\bibinfo {volume} {186}},\ \href
  {https://doi.org/10.1016/j.scriptamat.2020.04.042}
  {10.1016/j.scriptamat.2020.04.042} (\bibinfo {year} {2020})\BibitemShut
  {NoStop}%
\bibitem [{\citenamefont {Runnels}\ \emph {et~al.}(2021)\citenamefont
  {Runnels}, \citenamefont {Agrawal}, \citenamefont {Zhang},\ and\
  \citenamefont {Almgren}}]{Runnels2021}%
  \BibitemOpen
  \bibfield  {author} {\bibinfo {author} {\bibfnamefont {B.}~\bibnamefont
  {Runnels}}, \bibinfo {author} {\bibfnamefont {V.}~\bibnamefont {Agrawal}},
  \bibinfo {author} {\bibfnamefont {W.}~\bibnamefont {Zhang}},\ and\ \bibinfo
  {author} {\bibfnamefont {A.}~\bibnamefont {Almgren}},\ }\bibfield  {title}
  {\bibinfo {title} {Massively parallel finite difference elasticity using
  block-structured adaptive mesh refinement with a geometric multigrid
  solver},\ }\bibfield  {journal} {\bibinfo  {journal} {Journal of
  Computational Physics}\ }\textbf {\bibinfo {volume} {427}},\ \href
  {https://doi.org/10.1016/j.jcp.2020.110065} {10.1016/j.jcp.2020.110065}
  (\bibinfo {year} {2021})\BibitemShut {NoStop}%
\bibitem [{\citenamefont {Agrawal}\ and\ \citenamefont
  {Runnels}(2023)}]{Agrawal2023}%
  \BibitemOpen
  \bibfield  {author} {\bibinfo {author} {\bibfnamefont {V.}~\bibnamefont
  {Agrawal}}\ and\ \bibinfo {author} {\bibfnamefont {B.}~\bibnamefont
  {Runnels}},\ }\bibfield  {title} {\bibinfo {title} {Robust, strong form
  mechanics on an adaptive structured grid: efficiently solving
  variable-geometry near-singular problems with diffuse interfaces},\
  }\bibfield  {journal} {\bibinfo  {journal} {Computational Mechanics}\
  }\textbf {\bibinfo {volume} {72}},\ \href
  {https://doi.org/10.1007/s00466-023-02325-8} {10.1007/s00466-023-02325-8}
  (\bibinfo {year} {2023})\BibitemShut {NoStop}%
\bibitem [{\citenamefont {Runnels}()}]{alamo}%
  \BibitemOpen
  \bibfield  {author} {\bibinfo {author} {\bibfnamefont {B.}~\bibnamefont
  {Runnels}},\ }\href
  {https://doi.org/https://zenodo.org/doi/10.5281/zenodo.10381767,} {\bibinfo
  {title} {{Alamo: Massively parallel implicit/explicit solid
  mechanics}}}\BibitemShut {NoStop}%
\bibitem [{\citenamefont {the AMReX Development~Team}\ \emph
  {et~al.}(2024)\citenamefont {the AMReX Development~Team}, \citenamefont
  {Almgren}, \citenamefont {Beckner}, \citenamefont {Blaschke}, \citenamefont
  {Chan}, \citenamefont {Day}, \citenamefont {Friesen}, \citenamefont {Gott},
  \citenamefont {Graves}, \citenamefont {Huebl}, \citenamefont {Katz},
  \citenamefont {Myers}, \citenamefont {Nguyen}, \citenamefont {Nonaka},
  \citenamefont {Rosso}, \citenamefont {Sexton}, \citenamefont {Williams},
  \citenamefont {Zhang},\ and\ \citenamefont {Zingale}}]{amrex}%
  \BibitemOpen
  \bibfield  {author} {\bibinfo {author} {\bibnamefont {the AMReX
  Development~Team}}, \bibinfo {author} {\bibfnamefont {A.}~\bibnamefont
  {Almgren}}, \bibinfo {author} {\bibfnamefont {V.}~\bibnamefont {Beckner}},
  \bibinfo {author} {\bibfnamefont {J.}~\bibnamefont {Blaschke}}, \bibinfo
  {author} {\bibfnamefont {C.}~\bibnamefont {Chan}}, \bibinfo {author}
  {\bibfnamefont {M.}~\bibnamefont {Day}}, \bibinfo {author} {\bibfnamefont
  {B.}~\bibnamefont {Friesen}}, \bibinfo {author} {\bibfnamefont
  {K.}~\bibnamefont {Gott}}, \bibinfo {author} {\bibfnamefont {D.}~\bibnamefont
  {Graves}}, \bibinfo {author} {\bibfnamefont {A.}~\bibnamefont {Huebl}},
  \bibinfo {author} {\bibfnamefont {M.}~\bibnamefont {Katz}}, \bibinfo {author}
  {\bibfnamefont {A.}~\bibnamefont {Myers}}, \bibinfo {author} {\bibfnamefont
  {T.}~\bibnamefont {Nguyen}}, \bibinfo {author} {\bibfnamefont
  {A.}~\bibnamefont {Nonaka}}, \bibinfo {author} {\bibfnamefont
  {M.}~\bibnamefont {Rosso}}, \bibinfo {author} {\bibfnamefont
  {J.}~\bibnamefont {Sexton}}, \bibinfo {author} {\bibfnamefont
  {S.}~\bibnamefont {Williams}}, \bibinfo {author} {\bibfnamefont
  {W.}~\bibnamefont {Zhang}},\ and\ \bibinfo {author} {\bibfnamefont
  {M.}~\bibnamefont {Zingale}},\ }\href
  {https://doi.org/10.5281/zenodo.11099373} {\bibinfo {title}
  {Amrex-codes/amrex: Amrex 24.05}} (\bibinfo {year} {2024})\BibitemShut
  {NoStop}%
\bibitem [{\citenamefont {Schmidt}\ \emph {et~al.}(2022)\citenamefont
  {Schmidt}, \citenamefont {Quinlan},\ and\ \citenamefont
  {Runnels}}]{schmidt2022self}%
  \BibitemOpen
  \bibfield  {author} {\bibinfo {author} {\bibfnamefont {E.~M.}\ \bibnamefont
  {Schmidt}}, \bibinfo {author} {\bibfnamefont {J.~M.}\ \bibnamefont
  {Quinlan}},\ and\ \bibinfo {author} {\bibfnamefont {B.}~\bibnamefont
  {Runnels}},\ }\bibfield  {title} {\bibinfo {title} {Self-similar diffuse
  boundary method for phase boundary driven flow},\ }\href@noop {} {\bibfield
  {journal} {\bibinfo  {journal} {Physics of Fluids}\ } (\bibinfo {year}
  {2022})}\BibitemShut {NoStop}%
\end{thebibliography}%

\end{document}